\documentstyle[pra,eqsecnum,aps]{revtex}

\begin{document}
\draft

\title{Wavefunction Collapse and Random Walk}
\author{Brian Collett and Philip Pearle}
\address{Department of Physics, Hamilton College, Clinton, NY  13323}
\date{\today}

\maketitle

\begin{abstract}
{Wavefunction collapse models modify Schr\"odinger's equation so that it describes 
the rapid evolution of a superposition of macroscopically distinguishable states to one of them.   
This provides a phenomenological basis for a physical resolution to the so-called ``measurement problem." 
Such models have experimentally testable differences from standard quantum theory. The most well developed
such model at present is the Continuous Spontaneous Localization (CSL) model in which 
a fluctuating classical field interacts with particles to cause collapse. One ``side 
effect" of this interaction is that the field imparts energy to the particles: experimental 
evidence on this has led to restrictions on the parameters of the model, suggesting that 
the coupling of the classical field to the particles must be mass--proportional.   Another ``side 
effect," is that the field imparts momentum to particles, causing a small blob of matter to 
undergo random walk. Here we explore this in order to supply predictions which could be experimentally tested. 
We examine the translational diffusion of a sphere and a disc, 
and the rotational diffusion of a disc, according to CSL. 
For example, we find that the rms distance an isolated $10^{-5}$cm radius sphere 
diffuses is $\approx$ (its diameter, 5 cm) in (20 sec, a day), and that 
a disc of radius $2\cdot 10^{-5}$cm and thickness $.5\cdot 10^{-5}$cm diffuses 
through $2\pi$rad in about 70sec (this assumes the "standard" CSL 
parameter values). The comparable rms diffusions of standard quantum theory 
are smaller than these by a factor $10^{-3\pm 1}$.  It is shown that the CSL diffusion in air at STP is 
much reduced and, indeed, is swamped by the ordinary Brownian motion. It is
also shown that the sphere's diffusion in a thermal radiation bath at room temperature is comparable 
to the CSL diffusion, but is utterly negligible at liquid He temperature. Thus, in order to 
observe CSL diffusion, the pressure and temperature must be low.  At the low reported pressure 
of $<5\cdot10^{-17}$Torr, achieved at $4.2^{\circ}$K, the mean time between
air molecule collisions with the (sphere, disc) is $\approx$(80, 45)min.  This is ample time for 
observation of the putative CSL diffusion with the standard parameters and, it is pointed out, 
with any parameters in the range over which the theory may be considered viable.  This encourages 
consideration of how such an experiment may actually be performed, and the paper closes 
with some thoughts on this subject.}  
\end{abstract}


\section{Introduction}\label{Section I}

	Schr\"odinger was troubled by the collapse postulate associated with Bohr's 
``Copenhagen" version of quantum theory. This requires a superposition of 
macroscopically distinguishable states (an ill-defined concept), upon observation 
(another ill-defined concept), to be suddenly replaced by one of those states. In his 
famous ``cat paradox" paper \cite{Schrodinger} Schr\"odinger wrote this ``is the most 
interesting part of the entire theory," saying that it prevented one from ascribing 
reality to the wavefunction ``because from the realism point of view observation is a 
natural process like any other and cannot {\it per se} bring about an 
interruption of the orderly flow of events."  Dynamical 
wavefunction collapse models resolve Schr\"odinger's difficulty, allowing one 
to ascribe reality to the wavefunction (somewhat ironically) by altering 
Schr\"odinger's own equation, so that the collapse takes place in orderly 
and well--defined fashion.   

	The Continuous Spontaneous Localization (CSL) model \cite{PearleCSL,GPR},  
based upon previous models by Ghirardi, Rimini and Weber (GRW) \cite{GRW} and one of the authors 
\cite{Pearle} is the most well-developed collapse model at present \cite{others,PearleNaples}. In it, to 
Schr\"odinger's equation is added a term which contains a randomly fluctating classical field 
$w(\bf x ,t)$ that interacts with particles, bringing about collapse.  

	Although collapse is the desired and main effect, there are also ``side effects."  
Because collapse narrows wavefunctions, particles gain energy from the field in this process 
\cite{GR,Squires,Ballentine}.  Experimental tests\cite{PearleSquires,Collett,Ring}  
have resulted in restrictions on the 
range of permissable parameters for the model, sugggesting that the coupling 
between the field and particles (which determines the particle's collapse rate) 
is proportional to particle mass: thus, for a material object undergoing collapse, its nucleons are 
much more responsible for this behavior than are its electrons. In this paper we discuss another side effect:
the random impulses particles get from the field results in random walk of objects.  Indeed, in one 
of the earliest attempts at a dynamical collapse model, Karolyhazy \cite{Karolyhazy} discussed such behavior. 
Here we discuss it in the context of the CSL model, in order to see if the effect is measureable.   
 
	We first consider a sphere undergoing translational random walk.  
Section II summarizes the needed formalism associated with the usual Brownian motion in both 
air at temperature T and in a radiation bath at temperature T.  Section III summarizes the results 
(of calculations given in the appendices, as are most of the detailed calculations in this paper)
associated with CSL--induced random walk of the sphere. 
Section IV puts numerical values into these equations in three realms of air--sphere interaction: 
viscous, molecular and impact, in order of decreasing air molecule number density. It becomes clear that, 
in order to observe CSL diffusion, the air density must be low enough 
so that the mean time between air--sphere impacts is large compared to the 
time over which diffusion may be observed. 

	In Sections V and VI we turn to discuss 
respectively translational and rotational diffusion of a disc.   
Rotational diffusion of a disc will be the subject of our experimental proposal,     
because a small translation distance (e.g., the disc radius) becomes, when it is a distance of rotation, 
equivalent to a large fraction of $2\pi$rad and therefore 
more readily discernable. We consider a disc rather than a sphere  
because a perfect homogeneous sphere undergoes no CSL rotational diffusion 
since its rotated quantum states are identical. (An actual sphere's rotated states are slightly different so it 
does undergo a very small amount of collapse and rotational diffusion.) 
We find, for example, that a disc of radius $2\cdot 10^{-5}$cm and thickness $.5\cdot 10^{-5}$cm diffuses 
through $2\pi$rad in about 70sec: this assumes the "standard" values of the two  
parameters, proposed by GRW\cite{GRW} for their model and taken over into CSL, 
which may be characterized as the time $\lambda^{-1}=10^{16}$sec it takes an isolated 
nucleon in a superposition of two localized states separated 
by a distance greater than $a=10^{-5}$cm to collapse to one of those states. A 
pressure of $<5\cdot 10^{-17}$Torr is attainable\cite{Gabrielse} and, at this pressure, we 
find the mean collision time 
between air molecules and the disc is about 45 minutes. 
These results are so encouraging that we consider, in section VII, the full range 
of ($\lambda$, $a$) parameter values over which the theory may be considered viable (as well as 
the parameter proposal of Penrose\cite{others} based upon gravity).  This indicates that 
experiments to observe diffusion of small objects could provide a definitive test of 
CSL and other collapse models.  Therefore, in section VIII, we make a preliminary experimental proposal 
whose details we hope to examine in a future paper.  
 
\section{Brownian Motion Review}\label{Section II}

\subsection {Diffusion}\label{II A}

	It is useful to review the usual Brownian motion formalism \cite{Mazo}.  The Fokker-Planck 
equation for the probability density $\rho ({\bf x}, {\bf v}, t)$ for the position and 
velocity of the center of mass (CM) of a 
randomly walking sphere (radius $R$, mass $M$, 
density $D$) in a thermal bath at temperature $T$ is 

\begin{equation}\label{2.1} {\partial\rho\over\partial t}=
\sum_{j=1}^{3} \bigg\{ -v^{j}{\partial\rho\over\partial x^{j}}+{1\over\tau}{\partial v^{j}\rho\over\partial v^{j}}
+{\beta\over\tau^{2}}{\partial^{2}\rho\over\partial {v^{j}}^{2}} \bigg\}\end{equation}

\noindent where (as will be seen) $\tau$ characterizes the time to reach 
thermal equilibrium and $(2\beta t)^{1/2}$ 
is the equilibrium rms diffusion distance in time t. Eq. (2.1) can be used to calculate averages: 
${\bar f}(t)\equiv\int\int d{\bf x}d{\bf v}\rho ({\bf x}, {\bf v}, t)f({\bf x}, {\bf v})$.  
By multiplying Eq. (2.1) by $x^{j}$ and integrating by parts, and likewise for $v^{j}$, we obtain 
$d\overline {x^{j}}/dt=\overline {v^{j}}$, $d\overline {v^{j}}/dt=-\overline {v^{j}}/\tau$, so 
$\overline {v^{j}}=v^{j}(0)\exp -t/\tau$ and

\[\overline {x^{j}}=v^{j}(0)\tau \big[1-e^{-t/\tau}\big]\]

\noindent (assuming $x^{j}(0)=0$).  Likewise, $d\overline {{{x^{j}}^{2}}}/dt=2\overline {x^{j}v^{j}}$,  
$d\overline {x^{j}v^{j}}/dt=\overline {{{v^{j}}^{2}}}- \overline {x^{j}v^{j}}/\tau$, 
$d\overline {{{v^{j}}^{2}}}/dt=-2\overline {{{v^{j}}^{2}}}/\tau+2\beta/\tau^{2}$, so we obtain 
 
\begin{mathletters}
\label{all2.2}
\begin{equation}
\overline {{v^{j}}^{2}}-{\overline {v^{j}}}^{2}={\beta\over\tau}\big[1-e^{-2t/\tau}\big]
\end{equation}
\begin{equation}
(\Delta x)^{2}\equiv\overline {{{x^{j}}^{2}}}-{\overline {x^{j}}^{2}
=2\beta\tau}\bigg[t/\tau-\big(1-e^{-t/\tau}\big)-{1\over 2}{\big(1-e^{-t/\tau}\big)}^{2}\bigg]
\end{equation}
\begin{equation}
\thinspace (\Delta x)^{2}\thinspace\thinspace\thinspace\thinspace\thinspace\thinspace_
{\overrightarrow{t<<\tau}}\thinspace\thinspace\thinspace\thinspace\thinspace\thinspace
{2\beta t^{3}\over3\tau^{2}}[1-{t\over 4\tau}+...],\thinspace\thinspace\thinspace\thinspace\thinspace\thinspace
 \thinspace\thinspace\thinspace\thinspace\thinspace
 \thinspace(\Delta x)^{2}\thinspace\thinspace\thinspace\thinspace\thinspace\thinspace_{\overrightarrow{t>>\tau}}
 \thinspace\thinspace\thinspace\thinspace\thinspace\thinspace2\beta t.
\end{equation}
\end{mathletters}

\noindent We particularly call attention to the $\sim t^{3}$ diffusion for $t<<\tau$.

  One may readily express the variables $\beta$ and $\tau$ in terms of physical quantities. From 
$Md\overline {v^{j}}/dt=-M\overline {v^{j}}/\tau$ we see that the damping force is  
$-(M/\tau){\bf v}\equiv-\xi{\bf v}$.  According to the equipartition theorem, 
the equilibrium value of $\overline {{{v^{j}}^{2}}}$ is $kT/M$ so, 
by Eq. (2.2a), $kT/M=\beta/\tau$.  Thus

\begin{equation}\label{2.3}
\beta=kT/\xi, \thinspace\thinspace\thinspace\thinspace\thinspace\thinspace\tau=M/\xi.
\end{equation}

\subsection {Viscosity Factor}\label{II B}

	We now consider various expressions for $\xi$.  
In the case of a sphere in a fluid, as is well known, according to Stokes, the drag coefficient is 
 
\begin{equation}\label{2.4}
\xi=6\pi\eta R, \qquad (l_m <<R), 
\end{equation}

\noindent where $\eta$ is the viscosity of the fluid  and 
$l_m$ is the molecular mean free path. 

	If the fluid is a gas, as the gas density decreases there are three realms for air-sphere interaction.  
At high enough density so that $l_{m}<<R$, is the viscous realm: here $\xi$ is given by Stokes law (2.4).  At lower density, 
where $l_{m}>>R$, is the molecular realm. Here a colliding molecule may be 
considered to have a thermal velocity distribution since its last collision before hitting the 
sphere occurs so far away from the sphere that it is unaffected by the sphere's velocity (i.e., 
viscosity and hydrodynamic considerations are irrelevant). However, in this case, many molecular collisions 
occur over the shortest resolvable time
interval so the Brownian motion assumptions still apply, and $\xi$ is given by Eq. (2.5) below. 
At very low density, over time intervals 
where individual molecular collisions with the sphere 
can be resolved, is the impact realm where the Brownian motion assumptions no longer apply.

	In the molecular realm, Stokes law is no longer accurate. Experimental 
investigations into the correction to Stokes law, 
begun by Millikan\cite{Millikan} 
and continued to this day, are fitted by 

\[\xi=\frac{6\pi\eta R}{1+(l_{m}/R)[\alpha+\beta\exp-(\gamma R/l_{m})]}
\]

\noindent where e.g., recent measurements\cite{Aerosol} on polystyrene spheres give 
$\alpha\approx1$, $\beta\approx.6$ and $\gamma\approx1$.  In the limit $l_{m}>>R$ it is readily 
shown\cite{Cunningham,Epstein}, assuming specular reflection of air molecules (other assumptions moderately  
alter the numerical coefficients), that $\alpha=3/2$, $\beta=0$. In this case, using $\eta=(1/3)nm_{g}{\overline u}l_{m}$ 
($n$ is the gas molecular number density, $m_{g}$ is the mass of a gas molecule and  ${\overline u}$ is its mean velocity) 
in the above equation, the dependence upon $l_{m}$ disappears as one expects, resulting in 

\begin{equation}\label{2.5}
\xi=(4\pi/3)nm_{g}{\overline u}R^{2}=(8/3)nR^{2}(2\pi m_{g}kT)^{1/2}, \qquad (l_m >>R)  
\end{equation}
 
\noindent where we have used ${\overline u}=(8kT/\pi m_{g})^{1/2}$.

	In the case where the sphere moves in thermal 
radiation it is shown in Appendix D that a result of Einstein and Hopf\cite{EinsteinandHopf,Einstein} 
may be adapted to obtain, for a dielectric sphere of large dielectric constant (but also true up to a 
numerical constant for other shaped objects, where $R^{3}$ is replaced by the object's volume),

\begin{equation}\label{2.6}
\xi={4(2\pi)^{7}\over 135}\hbar R^{6}\bigg({kT\over\hbar c}\bigg)^{8}.
\end{equation}

	When Brownian motion assumptions apply, regardless of the physical source of $\xi$, 
the rms diffusion distance $\Delta x$'s long time and short time behaviors differ. 
Einstein's well known result\cite{Einstein1904} for $t>>\tau$ and the result for 
$t<<\tau$ follow from Eqs. (2.2c), (2.3):
	
\begin{mathletters}
\label{alll2.7}
\begin{equation}	
\Delta x\quad _{ \overrightarrow{t>>\tau}}\quad \bigg[{2kTt\over\xi}\bigg]^{1/2}
\end{equation}	
\begin{equation}
\Delta x\quad _{ \overrightarrow{t<<\tau}}\quad \bigg[{2kT\xi t^{3}\over3M^{2}}\bigg]^{1/2}.
\end{equation}
\end{mathletters}

\section{CSL Random Walk of a Sphere}\label{Section III}

\subsection {Diffusion of Center of Mass}\label{III A}

	In the case of CSL, we consider an ensemble of sphere CM wavefunctions $\langle{\bf q}|\psi,t\rangle_{w}$, 
each evolving under a particular sample field $w(\bf x ,t)$.  They 
are described by the density matrix $\rho (t)$ whose evolution equation (see Appendix A) satisfies

\begin{eqnarray}\label{3.1}
&&{\partial \langle{\bf q}|\rho(t)|{\bf q}'\rangle \over \partial t}= 
-i \langle{\bf q}|\bigg[ {{\bf P}^{2}\over 2M}, \rho(t)\bigg]|{\bf q}'\rangle\nonumber\\   
&&\qquad\qquad\qquad\qquad\qquad -{\lambda N^{2}\over V^{2}}\int\int_{V} 
d{\bf z}d{\bf z}'\bigg[\Phi({\bf z}-{\bf z}')-\Phi({\bf z}-{\bf z}'+{\bf q}-{\bf q}')\bigg]
\langle{\bf q}|\rho(t)|{\bf q}'\rangle 
\end{eqnarray}

\noindent under the approximation that the mass in the sphere is uniformly spread throughout it.  
In Eq. (3.1), ${\bf P}$ is the CM momentum operator, 
$V$ is the volume of the sphere and $\Phi({\bf z})\equiv \exp - {\bf z}^{2}/4a^{2}$.  
The electrons have been neglected because of their smaller mass and lower collapse rate, and 
the proton and neutron masses are taken to be equal for simplicity, so Eq. (3.1) 
depends just upon the nucleon number $N$. 

	Eq. (3.1) may be used to calculate ensemble averages of expectation values of operators: 
$\overline{\langle F\rangle}(t)\equiv Tr[F\rho(t)]$. The trace of the CSL 
term in Eq. (3.1) multiplied by the CM position operators 
$Q^{j}$ (or any function of them), by $P^{j}$ or by $Q^{j}P^{j}+P^{j}Q^{j}$ vanishes.  Thus we obtain 
$d\overline {\langle Q^{j}\rangle}/dt=\overline {\langle P^{j}\rangle}/M$,  
$d\overline {\langle P^{j}\rangle}/dt=0$ and so $\overline{\langle Q^{j}\rangle}(t)=0$ 
(assuming $\langle Q^{j}\rangle (0)=0$) and $\overline {\langle P^{j}\rangle}(t)=0$ (assuming $\langle P^{j}\rangle (0)=0$).
  
	However, the trace of the CSL term in Eq. (3.1) multiplied by 
${P^{j}}^{2}$ does not vanish.  Collapse increases 
energy because it narrows wavefunctions and the references given in section 1 show that (neglecting the  
collapse behavior associated with the electrons) the rate of energy increase is given by 

\begin{equation}\label{3.2}
{d\over dt}\overline{\langle H\rangle}={3\lambda\hbar^{2}  N^{2}\over 4M a^{2}},
\end{equation}

\noindent  As is shown in Appendix A, for the CM part of the energy it follows from Eq. (3.1) that 

\begin{equation}\label{3.3}
{d\over dt}{\overline {{{\langle {P^{j}}^{2}\rangle}}}\over 2M}={\lambda\hbar^{2} N^{2}f(R/a)\over 4M a^{2}}.  
\end{equation}

\noindent The factor $f$ essentially characterizes the collapse rate 
when the sphere is displaced by the distance $a$ (see the discussion after Eq. (A10) in Appendix A). 
$f(R/a)$, given in analytic form in Eq. (A9b), is a monotonically decreasing function of its argument, with $f(0)=1$, 
$f(1)=.62$ and $f(R/a)\rightarrow 6(a/R)^{4}$ for $R>>a$.  Summing Eq. (3.3) over the 
three values of $j$ and comparison with Eq. (3.2) shows that, for small $R/a$, the 
excitation of the CM accounts for almost all of the sphere's energy increase but, as $R/a$ increases,
internal nuclear excitation (too small to observe at present) accounts for more of it. 

	Therefore, using Eq. (3.1), 
since $d\overline {{{\langle {Q^{j}}^{2}\rangle}}}/dt=\overline {\langle Q^{j}P^{j}+P^{j}Q^{j}\rangle}/M$, 
$d \overline {\langle Q^{j}P^{j}+P^{j}Q^{j}\rangle}/dt=2\overline {{{\langle {P^{j}}^{2}\rangle}}}/M$ and 
$d\overline {{{\langle {P^{j}}^{2}\rangle}}}/dt$ is given by Eq. (3.3), we find 

\begin{equation}\label{3.4}
 \overline {{{\langle {Q^{j}}^{2}\rangle}}} =
 \langle \bigg( Q^{j}+\frac{P^{j}t}{M}\bigg)^{2}\rangle (0)+{\lambda\hbar^{2}f(R/a)t^{3}\over 6 m^{2}a^{2}} 
\end{equation}
	
\noindent where $m$ is the mass of a nucleon.  
We note the $\sim t^{3}$ diffusion associated with a random force without damping.     
This occurs essentially because the average square velocity is 
increasing so the distance of each ``step" in the random walk increases with time.  
In Eq. (3.4) we have utilized $M=Nm$ to emphasize that, for $R<<a$, the 
diffusion is ``universal," i.e., independent of the material and size (or, it turns out, shape) of the 
piece of matter and, in general, that the dependence on $N$ is only indirect, through the sphere's radius R. 

\subsection {Wavepacket Width}\label{III B}

	Eq. (3.4) is the result needed to describe CSL random walk.  However, it is necessary to show that the $\sim t^{3}$ 
term is not due to an increase in the width of the CM wavepackets in the ensemble but truly due to the 
diffusion of the centers of the packets. That is, the ensemble mean square wavepacket width is 
$\overline{s^{2}}\equiv\overline{{{\langle [Q^{j}-\langle Q^{j}\rangle]^{2}\rangle}}}$, 
what we want is the ensemble mean of the squared  
displacement of the center of the wavepackets $\overline{\langle Q^{j}\rangle^{2}}$ 
and what we've got from Eq. (3.4) is $\overline {{\langle {Q^{j}}^{2}\rangle}}=\overline{s^{2}}
+ \overline{\langle Q^{j}\rangle^{2}}$. 

	Under the combined influence of the collapse (which tends to narrow wavefunctions) and the 
normal Schr\"odinger evolution of a free object (which tends to expand wavefunctions), 
$\overline{s^{2}}$ tends to an equilibrium size in a characteristic time $\tau_{s}$.   
This has been discussed in the context of the GRW 
model\cite{GRW,BGG,BG} and for a particle in a simple continuous collapse model by Diosi\cite{Diosi4}.  We discuss 
it for the CSL model in Appendix B. It requires a separate treatment  
because $\overline{\langle Q^{j}\rangle^{2}}$ (which is 
{\it not} ${\overline{\langle Q^{j}\rangle}}^{2}=0$) and so $\overline{s^{2}}$ cannot be found 
from the density matrix since they involve an ensemble average over a quantity that is quartic in the statevector.
  
	According to Appendix B, the asymptotic CM wavepacket width is the same for every wavepacket 
(i.e., no ensemble average need be involved):

\begin{equation}\label{3.5}
s^{2}(t)\thinspace\thinspace\thinspace\thinspace\thinspace\thinspace_{\overrightarrow{t>>\tau_{s}}} 
\thinspace\thinspace\thinspace\thinspace\thinspace\thinspace
s^{2}_{\infty}=\bigg[{a^{2}\hbar\over 2\lambda m N^{3} f(a/R)}\bigg]^{1/2} 
\end{equation}	

\noindent This expression for $s_{\infty}$ can be understood as follows.  If a wavepacket has width $s$, 
due to its Schr\"odinger evolution it expands a distance $\sim (\hbar/Ms)\Delta t$ in time $\Delta t$. 
Due to the collapse evolution it contracts a distance $\sim$ (collapse rate)$\Delta t\cdot$(fractional decrease)$s$. 
As discussed in Appendix A (after Eq. (10)) the collapse rate is $\lambda N^{2}f$.  The fractional decrease is 
that which would occur if a gaussian of width $s<<a$ is multiplied by a gaussian of width $a$, namely 
$(s/a)^{2}$.  Thus it contracts a distance $\lambda N^{2}f(s/a)^{2}s\Delta t$.  Equating the 
Schr\"odinger expansion to the collapse contraction and solving for $s^{2}$ gives the result $\sim$(3.5).  

	The characteristic time to reach this width is 

\begin{equation}\label{3.6}
\tau_{s}={Nms^{2}_{\infty}\over\hbar}.
\end{equation}

\noindent This can be understood as the time it takes a packet of width $s_{\infty}$ to expand by 
the distance $s_{\infty}$: $(\hbar/Ms_{\infty})\tau_{s}=s_{\infty}$.   

	Appendix B also shows that the CSL diffusive behavior soon becomes the dominant contribution to 
$\overline{\langle {Q^{j}}^{2}\rangle}$ once equilibrium has been reached: 
with initial equilibrium conditions, Eq. (3.4) becomes 

\begin{equation}\label{3.7}
\overline{\langle {Q^{j}}^{2}\rangle}=s_{\infty}^{2}+s_{\infty}^{2}\bigg[{t\over \tau_{s}}+
{t^{2}\over 2\tau_{s}^{2}}+{t^{3}\over 12\tau_{s}^{3}}\bigg]
\end{equation}

\noindent Eqs. (3.5), (3.6) are derived in Appendix B under the assumption that $s^{2}(t)<<a^{2}$.  
From Eq. (3.5) this implies $N>3\cdot 10^{7}$ nucleons which, for ordinary matter densities 
(1gm/cc $< D <$ 20gm/cc) 
means that Eqs. (3.5), (3.6) hold for $R>2\cdot 10^{-6}$cm.

\section{Translational Diffusion Of A Sphere: Numerical Values}\label{Section IV}

	We shall now put numbers into these equations so as to consider the conditions 
necessary to observe CSL-induced diffusion of a sphere. In the following we shall use  
the GRW parameter values, $\lambda^{-1}=10^{16}$sec and $a=10^{-5}$cm, until section VII 
when we consider the full range of allowable parameter values for the theory.

\subsection{CSL Diffusion Alone}\label{IV A}

	According to Eq. (3.4), under the CSL mechanism acting alone, the rms distance along an axis the sphere diffuses is 

\begin{equation}\label{4.1}
\Delta Q^{j}={\hbar\over m a}\bigg[{\lambda ft^{3}\over6}\bigg]^{1/2}
=6.5f^{1/2}{(t\thinspace\thinspace{\rm days})}^{3/2}\thinspace\thinspace{\rm cm}
\end{equation}

\noindent where $f=1$ for $R<<a$.  Since Eq. (4.1) does not depend directly upon $N$ 
it is independent of the density $D$.  It does 
depend upon $R$ through $f$, decreasing rapidly as $R$ increases:

\[ \Delta Q^{j}\thinspace\thinspace\thinspace_{\overrightarrow {R>>a}}
16(a/R)^{2}{(t\thinspace\thinspace{\rm days})}^{3/2}\thinspace\thinspace{\rm cm}.\]

\noindent Table 1 lists $\Delta Q$ for various values of $R$ and $t$.

	Although diffusion is all we shall be concerned with in the remainder of this paper, 
we give here some values for $s_{\infty}$ and 
$\tau_{s}$.  From Eqs. (3.5), (3.6) it follows that

\[ s_{\infty}\thinspace_{\overrightarrow{R<<a}}
3.8\cdot 10^{-7}D^{-3/4}(a/R)^{9/4}\thinspace{\rm cm}\thinspace,\quad  
s_{\infty}\thinspace_{\overrightarrow{R>>a}}
2.4\cdot 10^{-7}D^{-3/4}(a/R)^{5/4}\thinspace{\rm cm}\] 
\[\tau_{s}\thinspace_{\overrightarrow{R<<a}}.58 D^{-1/2}(a/R)^{3/2}\thinspace{\rm sec},\qquad
\tau_{s}\thinspace_{\overrightarrow{R>>a}} 
.23 D^{-1/2}(R/a)^{1/2}\thinspace{\rm sec}\]

\noindent ($D$ is in gm/cc).  Table 2 lists values 
of $s_{\infty}$ and $\tau_{s}$ for various values of $R$, 
for a sphere of density $D=1$gm/cc. (We note that  $\tau_{s}$ increases as $R$ moves away from $a$ 
in either direction since as $R$ decreases the collapse rate decreases and as $R$ increases the Schr\"odinger 
spreading rate decreases). 

	For example, an $R=10^{-5}$cm sphere's center of mass wavefunction reaches equilibrium size 
$s_{\infty}\approx 4\cdot 10^{-7}$cm in  $\tau_{s}\approx .6$sec, and diffuses $\Delta Q\approx$ (60microns, 5cm) 
in (1000sec, 1day). 

	It is worth examining the diffusion to be expected were the CSL hypothesis to 
be false ($\lambda=0$) and the Copenhagen concept of collapse 
somehow occurring ``upon observation" to be employed.  In this case we utilize 
Eq. (3.4), $\Delta Q_{QM}=\langle P/M\rangle(0)t$. 
If the sphere is observed at $t=0$, localized to $\approx 2R$, we may take $\langle P\rangle(0)\approx\hbar/4R$.  
If the sphere is then ``in the dark" (unobserved) until time $t$, we obtain from 
$\Delta Q_{QM}\approx t\hbar/[D(4/3)\pi R^{3}4R]$, for $R=10^{-5}$cm and $D=1$gm/cc, that 
$\Delta Q_{QM}\approx (6\cdot10^{-6}$cm, $5\cdot10^{-4}$cm) in (1000sec, 1day).  These numbers are 
smaller than their CSL counterparts by the respective factors $(10^{-3}, 10^{-4})$.

\subsection {Diffusion in Air}\label{IV B}

	The CSL diffusion distances in vacuum given in Table 1 are much reduced 
by the $-\xi {\bf v}$ damping due to collisions with air molecules.  

[Another effect of 
these collisions is that, as the air molecules 
collide with the sphere they become entangled with its states, increasing the effective collapse rate.   
This effect is complicated, 
depending upon the sphere's quantum state's differences of air molecule density 
in $a^{3}$ sized volumes surrounding the sphere. Because the air molecule density is much less than the 
sphere density and because the quantum states of the sphere which compete in the collapse ``game" are 
so spatially close, I shall ignore this effect.]  

	The Fokker-Planck equation for the combined CSL and Brownian diffusion in air, which replaces Eq. (2.1), is 	
 
\begin{equation}\label{4.2} {\partial\rho\over\partial t}=
\sum_{j=1}^{3} \bigg\{ -v^{j}{\partial\rho\over\partial x^{j}}+{\xi\over M}{\partial v^{j}\rho\over\partial v^{j}}
+\bigg[{kT\xi\over M^{2}}+
{\lambda \hbar ^{2}f\over 4 m^{2}a^{2}}\bigg]{\partial^{2}\rho\over\partial {v^{j}}^{2}} \bigg\}
\end{equation}
	
\noindent The long time and short time diffusion expressions which replace Eqs. (2.7a,b) are

\begin{mathletters}
\begin{equation}\label{4.3a}
(\Delta x)^{2}\thinspace\thinspace\thinspace_{\overrightarrow{t>>\tau}} \bigg[{2kT\over \xi}+
\bigg({M\over\xi}\bigg)^{2}{\lambda \hbar ^{2}f\over 2 m^{2}a^{2}}\bigg]t.
\end{equation}
\begin{equation}\label{4.3b}
(\Delta x)^{2}\thinspace\thinspace\thinspace_{\overrightarrow{t<<\tau}} \bigg[{2kT\xi\over 3M^{2}}+
{\lambda \hbar ^{2}f\over 6 m^{2}a^{2}}\bigg]t^{3}.
\end{equation}
\end{mathletters}

\subsubsection {Viscous Realm}\label{IV B1}
	
	First consider the viscous realm. At room temperature $T_{0}$ 
and atmospheric pressure $p_{0}$, the mean free path of 
air (N$_{2}$ or O$_{2}$) is $\l_{m}\approx .6\cdot 10^{-5}$cm.  For $\xi$ we use 
Stokes' law (2.4) or the corrected equation following it 
(needed for $R=10^{-5}$cm since then $l_{m}\approx R$, which amounts to a 
40\% decrease in $\xi$ if $\alpha=3/2$ and $\beta=0$): with $\eta\approx 2\cdot 10^{-4}$gm/cm-sec we have 

\[ \tau=M/\xi\approx (2\cdot 10^{-6}, \thinspace\thinspace 10^{4}){\rm sec\thinspace\thinspace for\thinspace\thinspace}
R=(10^{-5}, \thinspace\thinspace 1){\rm cm}\]

\noindent Then we may apply Eq. (4.3a) for $t>\tau$  
to obtain 

\begin{mathletters}
\begin{equation}\label{4.4a}
\Delta x _{BR}=\bigg[{2kT\over \xi}t\bigg]^{1/2}\approx(.6,\thinspace\thinspace 1.4\cdot 10^{-3})(t {\rm days})^{1/2} 
{\rm cm\thinspace\thinspace for\thinspace\thinspace}
R=(10^{-5}, \thinspace\thinspace 1){\rm cm} 
\end{equation}
\begin{equation}\label{4.4b}
\Delta x_{CSL}=\bigg({M\over\xi}\bigg)\bigg({\lambda \hbar^{2}f\over 2m^{2}a^{2}}t\bigg)^{1/2} 
\approx 3\cdot 10^{-11}D(t {\rm days})^{1/2} 
{\rm cm\thinspace\thinspace for\thinspace\thinspace}
R\geq 10^{-5}{\rm cm}  
\end{equation}
\end{mathletters}

\noindent (note that Eq. (4.4b) is independent of $R$).  For $t<\tau$, where Eq. (4.3b) applies, 
the ratio $\Delta x_{CSL}/\Delta x _{BR}$ is the same as that given in Eqs. (4.4).  
 
	Clearly the CSL diffusion 
is swamped by the Brownian diffusion in the viscous realm (especially since it 
is the sum of the squares of Eqs. (4.4) which add in Eq. (4.3a)).

\subsubsection {Molecular Realm}\label{IV B2}

	We next turn to the molecular realm where $l_{m}>>R$, which can 
be achieved by lowering the air density through lowering the air pressure $p$ (which we shall give 
in units of picoTorr: 1pT=$10^{-12}$T). We focus upon  
$R=10^{-5}$cm spheres since, if $R>>a$, $\Delta x_{CSL}$ decreases $\sim R^{-2}$ and so is less easily observed (and also 
$\Delta x _{BR}\sim R^{-2}$ so no relative advantage is gained by increasing $R$). Moreover, 
spheres of this size have been typical of observations of Brownian motion in air \cite{Millikan,Andrade}.  
From Eq. (2.5) we find the time to reach thermal equilibrium is

\[\tau=M/\xi\approx 2\cdot 10^{9}(T/T_{0})^{1/2}(p\thinspace {\rm pT})^{-1}{\rm sec}\]

\noindent Since $\tau$ is so long for picoTorr pressure or less, we apply Eq. (4.3b) for $t<<\tau$: 

\begin{mathletters}
\begin{equation}\label{4.5a}
\Delta x_{BR}=\bigg[{2kT\xi\over 3M^{2}}t^{3}\bigg]^{1/2}\approx 2\cdot 10^{-4}(p\thinspace 
{\rm pT})^{1/2}(T/T_{0})^{1/4}D^{-1}t^{3/2}{\rm cm}
\end{equation}
\begin{equation}\label{4.5b}
\Delta x_{CSL}=\bigg[{\lambda \hbar ^{2}f\over 6 m^{2}a^{2}}t^{3}\bigg]^{1/2}
\approx 2\cdot 10^{-7}t^{3/2}{\rm cm}  
\end{equation}

\end{mathletters}

\noindent ($t$ is in sec).  

	It follows from Eqs. (4.5a,b), even at liquid He temperature $T=4.2^{\circ}$K 
and for a dense sphere $D=10$gm/cc, that $\Delta x_{BR}\approx\Delta x_{CSL}$ 
requires the very low pressure of $p\approx 10^{-3}$pT.  But, at this low pressure, the Brownian 
assumption of many molecule-sphere collisions occurring in the shortest observable time interval  
no longer applies. Therefore, observation of CSL diffusion requires 
the air density to be low enough to be in the impact realm.   

\subsubsection {Impact Realm}\label{IV B3}

	Consider the mean time between molecule-sphere collisions.  The mean number 
of collisions/sec-area of air molecules is $(n\overline{u} /4)$ so the mean time between 
molecule-sphere collisions is $\tau_{c}=[(n\overline{u} /4)(4\pi R^{2})]^{-1}$.   
Moreover, the change in speed of the sphere due to one collision with an air molecule is 
$\Delta v\approx \overline{u}(m_{g}/M)$.  With $\overline{u}\approx 4.5\cdot 10^{4}(T/T_{0})^{1/2}$cm/sec 
and $n\approx 2.5\cdot 10^{19}(p/p_{0})(T_{0}/T)$cm$^{-3}$ we obtain, for an $R=10^{-5}$cm sphere,  

\begin{equation}\label{4.6}
\tau_{c}\approx 2(T/T_{0})^{1/2}(p\thinspace {\rm pT})^{-1}{\rm sec}, \quad 
\Delta v\approx 5\cdot 10^{-5}(T/T_{0})^{1/2}{\rm cm/sec}.
\end{equation}

\noindent 

	[Incidentally, we can understand the molecular realm's Brownian motion in terms of the 
impact realm motion if we consider the impact realm but for $t>>\tau_{c}$, 
so that many collisions have occurred in time $t$ and so 
Brownian motion considerations apply.   
Then $\Delta x_{BR}$ in Eq. (4.5a) may be written in terms of the quantities in Eq. (4.6),  
using $\xi$ taken from Eq. (2.5): 

\[\Delta x_{BR}\sim[(kT/M^{2})nR^{2}(m_{g}kT)^{1/2}t^{3}]^{1/2}\sim \Delta v t[t/\tau_{c}]^{1/2},\].

\noindent This says that the Brownian diffusion distance in time $t$ is the distance the sphere goes with the speed 
it gets from a single collision multiplied by the square root of the number of collisions 
(the expected fluctuation in the number of collisions)].

	Eq. (4.5b) (the same as Eq. (4.1) gives the CSL diffusion distance in the impact realm. 
In order to observe CSL diffusion over the largest distance, one wants $\tau_{c}$ to be as long as possible and so, 
By Eq. (4.6), one wants the lowest possible pressure.  
An experiment conducted at pressure$<5\cdot 10^{-17}$Torr at $4.2^{\circ}$K has been reported\cite{Gabrielse}.  
In these conditions, the mean collsion time is $\tau_{c}\approx 80$min.  In this time, according to Eq. (4.5b), 
$\Delta x_{CSL}\approx .7$mm.  This should be readily observable, so much that it encourages one to 
contemplate an experiment to test CSL over a wide range of parameter values (Section VIII).

\subsection {Diffusion in a Thermal Radiation Bath}\label{IV C}

	For completeness, we note that, even when one eliminates collisions of the sphere with air molecules 
over some sufficiently long time interval, there is still thermal radiation to supply 
damping and random impacts and thus induce Brownian motion. However, as we shall soon see, 
this is very small at liquid Helium temperature.

	The viscosity coefficient in the case of thermal radiation, obtained in Appendix D  
and cited in Eq. (2.6), has the numerical value   	

\begin{equation}\label{4.7}
\xi_{RAD}\approx 4\cdot 10^{-29}(R/10^{-5})^{6}(T/T_{0})^{8} {\rm \thinspace gm/sec}
\end{equation}

\noindent ($R$ is in cm). 

	It follows from this that the time (2.3) to reach thermal equilibrium is 
	
\begin{equation}\label{4.8}	
\tau_{RAD}=M/\xi_{RAD}\approx 10^{14}D(R/10^{-5})^{-3}(T/ T_{0})^{-8}{\rm sec}.
\end{equation}	
	
\noindent At room temperature or less, $\tau_{RAD}$ is so long that only the case of $t<<\tau_{RAD}$ is of interest. 
Then, using Eqs. (2.7b) and (2.6), 

\begin{equation}\label{4.9}	
\Delta x_{RAD}=\bigg[{2kT\xi_{RAD}t^{3}\over3M^{2}}\bigg]^{1/2}
\approx 8D^{-1}(T/ T_{0})^{9/2}(t/10^{5})^{3/2}{\rm cm}.
\end{equation}

\noindent (note that (4.9) is independent of $R$). According to Eq. (4.9), at room temperature a 
 $D=1$gm/cc sphere of any radius will diffuse 
$\approx 7$cm/day due to thermal radiation alone.  This is essentially equal in magnitude to the CSL 
diffusion for an $R=10^{-5}$cm sphere (which is 
unaffected by such a small damping coefficient). However, at $T=4.2^{\circ}$K we have 
$\Delta x_{RAD}\approx 4\cdot 10^{-8}$cm in a day which is utterly negligible. 

	Therefore, at liquid He temperature, which is needed to obtain the low 
pressure of the impact realm, we need not consider the random walk due to thermal radiation.  

\section {Translational Diffusion Of A Disc}\label{V} 

	Because rotation through $2\pi$rads may be more easily detected than a comparable translation, 
our experimental proposal (section VIII) is based upon observing rotational diffusion. However, since a uniform sphere 
displays no CSL rotational diffusion we consider a more asymmetrical object, a disc.    
In this section, for completeness, we discuss translational diffusion of a disc. In section VI we shall discuss 
rotational diffusion of a disc.

\subsection {Brownian Diffusion}\label{VA}      

	Consider a disc of radius $L$ and thickness $b$.  
For Brownian motion, the time dependence of the rms diffusion is given in section II in terms of 
$\xi$ (e.g., Eqs. (2.7)).  In the viscous realm, for an oblate spheroid $(x^{2}+y^{2})/L^{2}+z^{2}/(b/2)^{2}=1$ 
(close enough to a disc), Lamb\cite{Lamb} shows that, for $b<<L$, 

\begin {mathletters}\label{5.1}
\begin {equation}
\xi\approx 16\eta L \qquad ({\rm motion\thinspace\thinspace perpendicular\thinspace\thinspace to \thinspace\thinspace face})
\end {equation}
\begin {equation}
\xi\approx (32/3)\eta L \qquad ({\rm motion\thinspace\thinspace along\thinspace\thinspace edge}).
\end {equation}
\end{mathletters}

\noindent This is not qualitatively different from Stokes law (2.4) (with $R\approx L$).  

	In the molecular realm where $l_m>>(L,b)$, one may readily calculate, as in 
references \cite{Cunningham,Epstein}, assuming specular reflection of the molecules in the disc rest frame, 

\begin {mathletters}\label{5.2}
\begin {equation}
\xi=4n L^{2}(2\pi m_{g}kT)^{1/2}\qquad 
({\rm motion\thinspace\thinspace perpendicular\thinspace\thinspace to \thinspace\thinspace face})
\end {equation}
\begin {equation}
 \xi=2n Lb(2\pi m_{g}kT)^{1/2}\qquad ({\rm motion\thinspace\thinspace along\thinspace\thinspace edge}).
 \end {equation} 
\end{mathletters}

\noindent Eq. (5.2a) is not qualitatively different from Eq. (2.5) for a sphere (with $L\approx R$).   
As for Eq. (5.2b), decreasing $b$ to reduce $\xi$ for edgewise motion does not reduce  
$\Delta x_{BR}/\Delta x_{CSL}$ since, from Eqs. (4.3), $[\Delta x_{BR}/\Delta x_{CSL}]^{2}\sim\xi/M^{2}\sim b^{-1}$.

\subsection {CSL Diffusion}\label{VB} 

	For CSL, the time dependence of the rms diffusion is given in Appendix A, Eq. (A.10) 
(and copied in Eq. (3.4)). $f=1$ 
for a disc with all dimensions $<<a$, so in this case the disc's diffusion is no different from that of a sphere 
with $R<<a$, Eq. (4.1).  There is a difference for $(b/2a)^{2}<<1$ and $(L/2a)^{2}>>1$:   
 
 \begin {mathletters}\label{5.3}
\begin {equation}
f\rightarrow (2a/L)^{2}\qquad ({\rm motion\thinspace\thinspace perpendicular\thinspace\thinspace to \thinspace\thinspace face})
\end {equation}
\begin {equation}
f\rightarrow (4/\sqrt{\pi})(a/L)^{3}\qquad ({\rm motion\thinspace\thinspace along\thinspace\thinspace edge}).
 \end {equation} 
\end{mathletters} 

	Thus the (thin) disc diffusion decreases less with increasing size than does the sphere's diffusion,     
for which $f\rightarrow 6(a/R)^{4}$.  Thus, if a larger object is needed for greater visibility, 
 a larger radius disc gives greater diffusion 
than does a sphere of the same radius. However, the conclusion reached in section IV for a sphere holds as well 
for a disc: the impact realm is required to effectively remove Brownian motion in order to see CSL translational 
diffusion of a disc.

\section{Rotational Diffusion of a Disc}\label{VI} 

	Rotational Brownian motion was (naturally) first considered by Einstein\cite{Einstein Rot}.  
The Fokker-Planck equation for an object rotating about a fixed axis through angle $\theta$ 
with angular velocity $\omega$ is identical  in form to Eq. (2.1) with the replacements $v^{j}\rightarrow \omega$, 
$x^{j}\rightarrow \theta$.  For rotation, the viscous torque on a sphere is $-\xi_{ROT}\omega$ where
\cite{Einstein Rot,Lamb2}

\begin{equation}\label{6.1}
\xi_{ROT}=8\pi\eta R^{3},\qquad (l_m <<R),
\end{equation}

\noindent which replaces Stokes' law, Eq. (2.4).  Eq. (2.7a) is replaced by 

\begin{equation}\label{6.2}
\Delta\theta_{BR}\quad _{ \overrightarrow{t>>\tau_{ROT}}}\quad \bigg[{2kTt\over\xi_{ROT}}\bigg]^{1/2}
\end{equation}

\noindent where $\tau_{ROT}=I/\xi_{ROT}$ and $I=(2/5)MR^{2}$ is the sphere's moment of inertia. We see 
from Eqs. (6.1), (6.2) compared with Eqs. (2.4), (2.7a) 
that $\Delta\theta_{BR}\approx\Delta x_{BR}/R$. 

	But, the case of a sphere is of no use to us.  In the approximation we make, 
where the nuclear mass is uniformly spread out over the sphere, 
there is no difference between two rotated quantum states of the sphere so, according to CSL, 
there is no collapse and therefore no random rotational motion
(without this approximation there {\it is} collapse 
and random rotation but it is very slow). However, 
CSL random rotation does occur for a nonspherical object. 

\subsection{Brownian Rotational Diffusion}\label{VIA} 

Here we consider 
rotational diffusion of a disc ``on edge" (i.e., oriented with the face of the disc in a vertical plane), 
of radius $L$ and thickness $b<<L$, 
in the molecular and impact realms ($l_{m}>>L$). For a sphere in these realms, 
if the molecules make elastic collisions with the sphere, they do not transfer 
momentum parallel to the sphere face and so do not cause any torque ($\xi_{ROT}=0$).  However, for the disc, 
a straightforward calculation (as in \cite{Cunningham,Epstein}) yields the torque $=-\xi_{ROT}\omega$ 
about an axis passing through the edge and center, where 

\begin{equation}\label{6.3}
\xi_{ROT}=(4/\pi)nL^{4}(2\pi m_{g}kT)^{1/2}  
\end{equation}

\noindent and Eq. (2.7b) is replaced by 

\begin{equation}\label{6.4}
\Delta \theta_{BR}\quad _{ \overrightarrow{t<<\tau_{ROT}}}\quad \bigg[{2kT\xi_{ROT} t^{3}\over3I^{2}}\bigg]^{1/2}
\approx 80\frac{(p\thinspace{\rm pT})^{1/2}(T/T_{0})^{1/4}t^{3/2}}{(D{\rm gm/cc})
(b{\rm d\mu})
(L{\rm d\mu})^{2}}{\rm rads}
\end{equation}

\noindent ($I\approx ML^{2}/4$). Here we have employed the rather weird unit $1{\rm d}\mu=10^{-5}$cm 
because the dimensions of the disc we are considering are such that the factors $(b{\rm d\mu})$, 
$(L{\rm d\mu})$ are of the order of unity. We only give Eq. (6.4), valid for $t<<\tau_{ROT}$, because   
the time to reach thermal equilibrium is so long at picoTorr pressures or less: 
$\tau_{ROT}=I/\xi_{ROT}\approx 5\cdot 10^{7} 
D(b{\rm \thinspace\thinspace in\thinspace\thinspace d\mu})(p{\rm pT})^{-1}(T/T_{0})^{1/2}$sec. 

	According to Eqs. (6.4) and (6.5) (below), Brownian motion dominates CSL diffusion even at 1pT pressure, for 
discs with dimensions of the order of $a=1d\mu$.  Therefore one must go to lower pressure, to the impact realm, to see 
CSL rotational diffusion.   

\subsection{CSL Rotational Diffusion}\label{VIB} 

	For CSL rotational diffusion, it is shown in Appendix C that 

\begin{equation}\label{6.5}
\Delta \theta_{CSL}\approx \frac{\hbar}{ma^{2}}\bigg(\frac{\lambda t^{3}f_{ROT}}{12}\bigg)^{1/2}
\approx .018f^{1/2}_{ROT}t^{3/2}{\rm rad}   
\end{equation}

\noindent where FIG. 1 contains a graph of $f_{ROT}(\alpha, \beta)$ vs. $\alpha\equiv (L/2a)$ 
for various values of $\beta \equiv (b/2a)$. For example, 
$f_{ROT}\approx 1/3$ for $b\approx .5a$ and $L\approx 2a$. 

	For this example, 
according to Eq. (6.5), $\Delta \theta_{CSL}$ 
diffuses through $2\pi$rad in about 70sec. If $\lambda$ were $10^{-4}$ times smaller i.e., 
$\lambda\approx 10^{-20}$  (and still $a=10^{-5}$cm), which is at the edge of where one may consider the theory to 
be viable (section VII, Eq. (7.3)), this time is about 25 min. 

	It is worth examining the rotational diffusion to be expected from 
standard quantum theory (as was done for translational diffusion at the end of section IVA).  From Eq. (C6) 
with $\lambda=0$ we have 

\[\Delta \theta_{QM}=\langle{\cal L}\rangle (0)t/I\]

\noindent where $\langle{\cal L}\rangle (0)$ is the expectation value of 
the angular momentum operator in the initial state. If the disc is observed at $t=0$, localized to $\Delta \theta\approx\pi /4$, 
then we may take $\langle{\cal L}\rangle (0)\approx \hbar/[2(\pi/4)]=2\hbar/ \pi$.  If the disc is ``in the dark" 
(unobserved) until time t, using $I=(Db\pi L^{2})(L^{2}/4)$, we obtain
 
\begin{equation}\label{6.6}
\Delta \theta_{QM}\approx \frac{8\hbar t}{\pi^{2}DbL^{4}}.
\end{equation}

\noindent With the choices $D=1$gm/cc, $b=.5\cdot10^{-5}$cm, $L=2\cdot10^{-5}$cm, we get 
$\Delta \theta_{QM}\approx 10^{-3}t$rad.  Thus, $\Delta \theta_{QM}\approx$(.1, 1, 86)rad in 
$t=$(100sec, 1000sec, 1day).  These numbers are smaller than their CSL counterparts 
by the respective factors (100, 300, 3000).  However, were $\lambda$ sufficiently small, 
this diffusion could be observed in the experiment we propose (section VIII).

\subsection{Gas-Disc Collisions}\label{VIC}

	The times given in the example of the previous section for diffusion 
through $2\pi$rad ($\approx70$sec for $\lambda^{-1}=10^{16}$sec, 
$\approx$25min for $\lambda^{-1}=10^{20}$sec) should be compared with the 
mean time between collisions of air molecules with the disc.
	  
	Assume Nitrogen molecular gas at temperature $4.2^{\circ}$K and pressure $5\cdot10^{-17}$Torr.  
The mean molecular speed is ${\overline u}=[8kT/\pi m_{g}]^{1/2}\approx 5.6\cdot10^{3}$cm/sec. 
The molecular density is $\rho=p/kT\approx115$particles/cc. The molecular flux is 
$J=\rho{\overline u}/4\approx 1.5\cdot10^{5}$particles/cm$^{2}$-sec.  

	The mean time between collisions (we consider collisions with the 2 faces of the disc 
but neglect collisions with the edge) is thus $\tau_{c}=1/(2J\pi L^{2})\approx45$min.

	We conclude this subsection with an estimate of the effect of a collision.  A Nitrogen molecule with speed $\overline u$ 
impacting perpendicular to the disc 
face at distance $L$ from the rotation axis ("worst possible case")  conveys to 
the disc an angular velocity 

\begin{equation}\label{6.7}
\omega=\frac{m_{g}{\overline u}L}{I}\approx \frac{33}{D(b{\rm \thinspace\thinspace}d\mu)
(L{\rm \thinspace\thinspace}d\mu)^{2}}{\rm rad/sec}
\end{equation}

\noindent  For the example we have been considering, this is $\omega\approx$ 8rad/sec.  Such a sudden 
jump in the angular velocity should be readily observable and distinguishable from the 
expected CSL behavior.  
  	
\section {Parameter Values}\label{Section VII}

	 In the previous sections of this paper, for clarity's sake, in 
numerical calculations we have used the 
values of the CSL parameters $(\lambda^{-1},a)$ suggested by GRW, namely $(10^{16}$sec$,10^{-5}$cm). 
However, these values have no theoretical underpinning and were simply chosen to give reasonable 
results: other values are possible. However, not all values are possible. 
  
	Therefore we examine already existing experimental and theoretical constraints on these parameters.
Any new experiment must be considered as placing further constraints. 
While one may hope that experiment reveals an "anomalous" random walk  
confirming the existence of a CSL-type collapse process and disclosing the values of the parameters, one 
should consider the possibility that this does not happen. We consider 
the additional constraints negative results could provide.   
In particular, we consider what would be needed to eliminate CSL as a viable resolution of the ``measurement problem".  
 
	We also discuss two other topics.  One is the random walk associated with a suggestion by Penrose of 
a connection between gravity and collapse. This can essentially be interpreted as giving the results of 
this paper with a particular value for $\lambda$. We also show the range of parameter values 
consistent with a speculation that the fluctuating field $w$ has a thermal basis, based upon 
cosmological considerations and an analogy between 
standard random walk and CSL random walk.

	In what follows, $\sim$ means up to a numerical factor not too far from 1. 

	An experiment which looks for photons emitted by the atoms in an underground shielded slug of Germanium  
places a limit on the number of bound electrons or 
nucleons "spontaneously" excited in Ge atoms\cite{Collett,Ring}. Spontaneous excitation of bound states is 
expected from the CSL collapse mechanism,  
which narrows electron and nucleon wavefunctions thereby giving these particles increased energy (presumably the energy 
for this comes from the fluctuating collapse-causing field $w$\cite{Pearleenergy}). For example, a 1s 
electron ejected from an Ge atom will result in radiation of an 11.1 keV (equal to its binding energy) 
shower of photons from the atom's remaining electrons as they cascade downward added to radiation equal to 
the kinetic energy of the ousted electron which it rapidly loses in collisions with other atoms. 
  
	The present experimental upper limit on the rate of photon pulses appearing in 1 KeV bins above 11 keV 
is $\approx$ .05 pulses/(keV kg day) \cite{AvignoneRing}.  
The theoretical excitation rate is conveniently expanded in a power series in (size of bound state/$a)^{2}$.  The 
first term in this series turns out to vanish identically if the collapse coupling constant 
is mass proportional \cite{PearleSquires}.  We have assumed this in the present paper 
(e.g., see Eq. (A.1) et. seq.) because, for atomic spontaneous excitation, the numerical coefficient of this first term 
is large enough to make the experiment sensitive to the relative coupling constant size of electrons and nucleons, and the 
results make mass-proportionality likely. The experiment is less sensitive to the second term in 
the series but the data on excitation rate of nucleons 
still provides a constraint (because of the now-assumed small electron coupling 
constant, the electron excitation rate data does not provide as strong a constraint). The theory gives  
probability/sec$\sim\lambda$(nucleon diameter/$a)^{4}$.  Since, in Ge, there 
are $8.3\cdot 10^{24}$atoms/kg and $A\approx 72$ and, using nuclear radius$\approx 1.4\cdot 10^{-13}A^{1/3}$cm and 
$8.6\cdot 10^{4}$sec/day, we get

\[ \lambda^{-1}a^{4}>2\cdot 10^{-15}. \]

	However, the strongest present constraint, based upon the same experimental data, was provided by Fu\cite{QFu}. 
He calculated the rate of radiation by a free electron (mass $m_{e}$) 
due to being shaken by the collapse mechanism.  He obtained 
for the number of photons of energy E radiated by an electron per second per energy the expression

\[ R(E)=\frac{\lambda (m_{e}/m)^{2}e^{2}\hbar}{4\pi^{2}a^{2}m^{2}c^{3}E}=
8.1\cdot10^{-38}\frac{(\lambda/a^{2})}{(\lambda/a^{2})_{GRW}}
\bigg(\frac{1}{E {\rm keV}}\bigg){\rm counts/(sec\thinspace\thinspace keV)}\rightarrow
2.1\cdot10^{-8}{\rm counts/(keV\thinspace\thinspace kg\thinspace\thinspace day)}.\]

\noindent The last term on the righthand side of the above equation  
gives the rate of radiation from the 4 valence electrons (essentially free) from each atom in 
a slug of Ge (using $8.29\cdot10^{24}$ atoms/kg for Ge) at $\approx$ 11 keV with the GRW parameter values. 
This and the experimental upper limit quoted above leads to the experimental constraint:
  
\begin{equation}\label{7.1}
\lambda^{-1}a^{2}>.4
\end{equation}

\noindent This constraint (labelled line 1) is graphed in Fig. 1: the allowed region is to the right of the line. 

	Diffusion experiments should do much better than (7.1) in constraining the parameter values.  For example, consider 
a rotational diffusion experiment such as we sketch in the next section.  One expects to be able 
to detect a $\Delta\Theta\approx \pi/2$ diffusion in 45 minutes.  If such a diffusion were not 
detected, Eq. (6.5) (where $\Delta\Theta_{CSL}$ goes as $\lambda^{1/2}/a^{2}$) gives  
			
\begin{equation}\label{7.2}
\lambda^{-1}a^{4}>10^{2}
\end{equation} 

\noindent (region to the right of the line labelled 2 in Fig.1). This amounts to being able to detect $\Delta\Theta_{CSL}$ 
a factor $10^{-3}$ times smaller than that expected using the GRW parameter values. 

	There is also what may be called a ``theoretical constraint"\cite{Collett,Ring}, although it is fairly rough.  The 
purpose of a collapse model is to account for the world as we see it.  The model may be considered to fail if it allows 
an observable object to remain in a superposition of two well-separated locations ``too long". How long is 
``too long"?  We might take that to be human perception time $\sim .1$sec. 

For a first example, 
consider an object which is just visible, a sphere of diameter $4\cdot 10^{-5}$cm, in a superposition 
involving a displacement $>>a$, with $a>4\cdot 10^{-5}$cm. The collapse time is 
$\sim\lambda^{-1}/N^{2}$, where $N$ is the number of particles in the sphere. If the sphere's density is 
$D\approx 1$gm/cc, then $N\approx 2\cdot 10^{10}$ and the condition $\lambda^{-1}/N^{2}<.1$sec implies 

\begin{equation}\label{7.3}
\lambda^{-1}<4\cdot 10^{19}
\end{equation} 

\noindent (region below the line labelled 3 in Fig.1). 

	For a second example, again consider the above sphere but with a superposition involving a 
displacement$<a$. In this case the collapse time is  
$\sim (4\lambda^{-1}a^{2})/[N\cdot$displacement$]^{2}$.  Using the smallest possible discernible displacement, 
$4\cdot 10^{-5}$cm, the condition that the collapse time is $<.1$sec implies

\begin{equation}\label{7.4}
\lambda^{-1}a^{2}<1.6\cdot 10^{10}
\end{equation} 

\noindent (region to the left of the line labelled 4 in Fig.1).

	These ``theoretical constrains" are rough but we may take them seriously enough to observe, 
from Fig. 1,  that constraints 
(7.2) and (7.4) still permit a narrow wedge-shaped range of allowed parameters. 

	However, suppose one were able to perform 
an experimental test of translational diffusion of a sphere with a precision for $\Delta Q$ 
that is $10^{-3}$ times smaller than $\Delta Q_{CSL}$ with the GRW parameters, and find a null result. 
According to Eq. (4.1) the resulting 
constraint is $\lambda^{1/2}/a<10^{-3}(10^{-16}$sec$^{-1})^{1/2}/(10^{-5}$cm$)$ or

\begin{equation}\label{7.5}
\lambda^{-1}a^{2}>10^{12}
\end{equation} 

It appears that the conflict between (7.5) and (7.4) would make CSL nonviable. 

	We close this section with two additional considerations. 
	
		First, in Appendix E we argue that a proposal by Penrose\cite{others},    
and other suggestions involving a gravitational basis for collapse\cite{Diosi,GGR,PearleSquires2},  
arrive at an effective value for $\lambda$: $\lambda_{G}\approx Gm^{2}/a\hbar\approx 2\cdot 10^{-23}$sec$^{-1}$  
when the object undergoing collapse is of size $\approx a$. With such a small value of $\lambda$,  
the ``theoretical constraint" inequality (7.3) is violated: for the superposed   
sphere states considered in obtaining (7.3), the collapse time is $\approx 10$sec, much longer than human perception time. 

	However, proponents of $\lambda=\lambda_{G}$ could argue for a weaker ``theoretical constraint"\cite{ABGG}. 
That is, when a human observer looks at the sphere, the detection process in the brain amounts to entangling the 
two spatially distinct sphere states with two spatially distinct states of brain particles. This 
extra entanglement, while only roughly estimable, appears to bring about collapse in less than human perception time.  

	Detection of diffusion with such a small value of $\lambda$ could be possible.  
For example, we may compare the expected rotational diffusion (6.6) of standard quantum theory 
with the expected CSL diffusion (6.5) with  $\lambda=\lambda_{G}$ and $a=10^{-5}$cm: $\Delta\Theta_{QM}\approx 10^{-3}t$rad  
and $\Delta\Theta_{G}\approx 10^{-5}t^{3/2}$rad.  These give, for times (45min, 3hour), 
$\Delta\Theta_{QM}\approx (2.7,$  10.8)rad and  $\Delta\Theta_{G}\approx (1.4,$ 11.2)rad.

	Last, we call the reader's attention to Appendix F, where we 
consider that the collapse-inducing fluctuations of $w$ may come from a thermal bath of some unspecified medium in 
thermal equilibrium with the $2.7^{\circ}$K cosmic radiation. The 
$\sim t^{3/2}$ time dependence of the CSL diffusion for a nucleon is identified with 
the standard Brownian motion at this temperature over an interval much less than the time it takes to reach thermal equilibrium. 
The latter time is taken to be $\gamma\cdot$(the age of the universe), with $\gamma>1$. We 
then obtain the equality (F.2), $\lambda^{-1}a^{2}\approx 10^{3}\gamma$, which is consistent with 
a wide range of parameter values, e.g., for $a=10^{-5}$cm this implies $\lambda^{-1}>10^{13}$sec.

\section {Experimental Considerations}\label{VII}

In order to observe quantum mechanical rotational diffusion, either that 
arising from standard quantum theory or from CSL, it is necessary to isolate a small 
object from all outside torques for a period of minutes to hours while measuring 
its rotational position. We believe that it is now possible to perform such an 
experiment by combining techniques from nanomachining and from atom/particle trapping.
Below, we shall consider the problems of creating suitable discs, suspending and isolating
them, removing their residual thermal energy, and monitoring their angular position
as a function of time.

Our first consideration is the production of suitable disc samples. 
Larger discs (2$\mu$m diameter) have already been fabricated from silicon dioxide using standard 
IC fabrication methods \cite{Chen} and recent work at the Cornell Nanofabrication Facility shows 
that it is now possible to create structures with lateral dimensions below 100nm and thicknesses less than 40nm 
\cite{Tannenbaum}. So it appears possible to make discs of suitable dimensions using current methods. 
As explained below, we suggest using discs of highly conducting metals such as copper or gold. 

We next consider suspending and isolating a disc for the duration of the experiment. 
We suggest utilizing a charged, conducting disc in a Paul trap \cite{Paul} 
in a very high vacuum. The Paul trap uses an alternating quadrupole 
electric field to suspend a charged particle. While the method was initially developed 
to confine atoms, it was quickly adapted to suspend larger objects. 
Wuerker at. al.\cite{Wuerker} injected small conducting microparticles using an electrostatic
method that also charged the particles in the injection process. Once
they had fed a cloud of particles into the trap they were able to select a single
particle to retain in the trap by manipulating the trap's operating fields.

More recently, Arnold and co-workers\cite{Arnold1} have used Paul traps and modified Paul traps to
confine single microparticles (a few $\mu$m in size) for optical experiments. One of their modified Paul traps
has been used to confine single microparticles to within the Brownian limit set by the
atmospheric gas in their traps. They add extra 
static electric fields to counter the effects of both gravity and
imperfections in the quadrupole shape of the main field\cite{Arnold2}. This leaves a perfectly force free
spot in the trap where the particle will sit, subject only to collisions with the gas. 
Pressures of less than $5\cdot 10^{-17}$ Torr have been reported \cite{Gabrielse}
in traps cooled with liquid He to 4$^{\circ}$K. As we have remarked in subsection VIC, 
for a disc of radius $2\cdot 10^{-5}$cm and thickness $.5\cdot 10^{-5}$cm, at these conditions 
the average interval between gas-disc
collisions is 45 minutes.  Moreover, the Poisson statistical nature of the collisions make it
likely to find intervals between collisions up to 90 minutes. This is quite long enough to observe
even the rotational diffusion predicted by standard quantum mechanics.
 
Although we are still engaged in studying the detailed dynamics of a charged conducting disc in the Paul
trap, it appears already that the positional trap also acts as an orientational trap 
and will suspend the disc ``vertically oriented" (with its flat surface parallel 
to the vertical symmetry axis of the field). 
There is no torque on a centrally positioned and vertically oriented disc 
causing it to rotate about the symmetry axis in the direction which we shall refer to as
the azimuthal direction. Thus the trap appears to suspend a charged disc in exactly the best
orientation to observe rotational diffusion in the azimuthal direction. What is currently under study is whether a 
displacement from center and/or tipping off vertical of the disc causes an azimuthal torque and, if so,  whether that 
should be minimized or not (i.e., if the extra motion is due to translational or rotational 
CSL diffusion, this might cause an increase of observable CSL-induced diffusive rotation). 
  
When first injected into the trap, the discs will possess considerable translational and
rotational kinetic energy. Arnold et. al. were able to remove this energy by the viscous
interactions with the gas in the cell. In a high vacuum experiment there is no gas to take
up this kinetic energy and a disc will continue to orbit the trap. However, if we
add a transverse magnetic field then the eddy currents set up in the disc will convert the
kinetic energy to thermal energy in the disc. The magnetic forces induced are proportional
to the velocity of the disc and so provide a true viscous force. A simple dimensional analysis
suggests that quite moderate fields, no more than a few kilogauss, will damp out the
mechanical energy in a few seconds, thus bringing the disc to rest at the null point of
the trap and vertically oriented.

Preliminary calculations show that light (e.g., from a laser) shone in along the symmetry 
axis of the trap will scatter from the disc in a pattern that exhibits azimuthal anisotropy: 
more light is scattered perpendicular to the faces of the disc than perpendicular to the edges.   
Thus the orientation of the disc about a vertical axis can be monitored by collecting the 
scattered light. It appears that the geometry of the Paul trap makes it particularly easy to 
collect light scattered from a particle at the center of the trap. 
If the electrodes are highly polished, then the geometry is such that photons scattered away 
from the symmetry axis will be funneled by the electrodes to emerge through the two gaps where the 
cap electrodes and the ring electrode do not meet. Moreover, the scattered photons will retain 
their azimuthal orientation so that light collected at the gaps will retain the azimuthal 
intensity distribution and so provide information about the orientation of the disc.
We suggest collecting the scattered photons with 8 photomultipliers operating as photon
counters spaced around each gap so that the disc orientation can be measured to within $45^{\circ}$.

The light which illuminates and scatters from the disc does so symmetrically and therefore 
exerts no average azimuthal torque on it. However, because the photons scatter 
randomly from the disc, they exert a random torque on it and so cause it 
to undergo diffusive rotation. Fortunately, this effect scales with the 
light intensity and thus can be minimized by using sufficiently weak 
illumination. Moreover, this illuminational diffusion can itself be measured in 
exactly the same way as any other rotational diffusion. Thus its effects 
can be eliminated by studying the behavior of the disc as the light 
level is reduced. The precise limit on the maximum amount of light that can be scattered without
materially affecting the precision of the experiment depends on the value of $\lambda$ that one
wishes to measure. For example, according to our present rough calculations, 
for the standard value of $\lambda = 10^{-16}$sec$^{-1}$, the disc
can scatter about 200 photons per second before the illuminational diffusion exceeds 5\% of the
CSL diffusion. A simple calculation shows that, if the disc were at rest, 
you would need to count photons for about 10 seconds
to localize it to within 45$^{\circ}$. Since CSL diffusion with that $\lambda$ should lead to
one revolution every 70 seconds the time resolution is quite adequate.

In order to observe diffusion for a lower value of $\lambda$, 
the maximum light level must be reduced accordingly. 
However, the lower value of $\lambda$ will lead to a slower rate of CSL diffusion and so allow
integration of light over a longer period. This makes up for the lower maximum light level and means
that the lower limit on $\lambda$ that can be measured is set by the vacuum and not by the
illumination. A 45-90 minute interval between gas-disc collisions sets a lower limit 
$\lambda <10^{-23}$sec$^{-1}$. This is low enough to allow the experiment to definitively test CSL and, 
if the CSL diffusion does not appear, to see the diffusion expected
from standard quantum mechanics.

\acknowledgments

We especially appreciate the help of Gordon Jones and we would also like to thank Frank Avignone,  
Peter Milloni, James Ring and Ann Silversmith for their contributions to this work.  
One of us (P.P.) would like to thank Harvey Brown and the Philosophy of Physics group at Oxford for 
the stimulating environment where the idea for this work originated.   

\appendix

\section {Translational diffusion In CSL}

	In CSL, the density matrix evolution of the wavefunction of a blob of matter containing $N$ particles, in the position 
representation $|{\bf x}_{1},...{\bf x}_{N}>\equiv |x>$, is given by\cite{PearleCSL,GPR}

\begin{equation}\label{A1} 
{\partial\over \partial t}\langle x|\rho (t)|x'\rangle =-i\langle x|[H,\rho (t)]|x'\rangle
-{\lambda\over 2}\sum_{i=1}^{N}\sum_{j=1}^{N}{m_{i}m_{j}\over m^{2}}
[\Phi({\bf x}_{i}-{\bf x}_{j})+\Phi({\bf x}'_{i}-{\bf x}'_{j})-2\Phi({\bf x}_{i}-{\bf x}'_{j})]
\langle x|\rho (t)|x'\rangle
\end{equation}

\noindent where

\begin{equation}\label{A2}
\Phi({\bf z})\equiv e^{-{\bf z}^{2}/ 4 a^{2}},
\end{equation}

\noindent $\lambda$ is the collapse rate for a proton and we have assumed mass-proportionality 
of the collapse coupling (see section VII), ${\bf x}_{j}$ is the position coordinate of the $j$th particle, 
$m_{j}$ is its mass, $m$ is the mass of the proton and $H$ is the
usual Hamiltonian.  In what follows we shall neglect the contribution of the electrons 
because of the smallness of the electron mass, and for simplicity take the mass of the neutron equal to $m$. 
Then, $N$ is the number of nucleons.  

	We wish to consider only the behavior of the center of mass (CM) of the blob. Accordingly, 
we trace Eq. (A1) over the relative coordinates ${\bf R}_{i}\equiv {\bf X}_{i}-{\bf Q}$ (eigenvalues ${\bf r}_{i}$) 
where ${\bf Q}\equiv \sum_{i}m_{i}{\bf X}_{i}/\sum_{i}m_{i} = N^{-1}\sum_{i}{\bf X}_{i}$ 
is the CM position operator (eigenvalues ${\bf q}$).
$\Phi({\bf x}_{i}-{\bf x}_{j})= \Phi({\bf r}_{i}-{\bf r}_{j})$ and $\Phi({\bf x}'_{i}-{\bf x}'_{j})=\Phi({\bf r}'_{i}-{\bf r}'_{j})$ 
are independent of ${\bf q}$ but  
$\Phi({\bf x}_{i}-{\bf x}'_{j})=\Phi({\bf r}_{i}-{\bf r}'_{j}+{\bf q}-{\bf q}')$.  
We shall also assume that the density matrix is the direct product of the internal and CM density matrices
(this neglects their entanglement due to the collapse-induced excitation of the internal nuclear states). 
The trace of Eq. (A1) over relative coordinates yields 

\begin{eqnarray}\label{A3} 
&&{\partial\over \partial t}\langle{\bf q}|\rho _{cm} (t)|{\bf q}'\rangle =
-i\langle{\bf q}|\bigg[ {P^{2}\over 2M},\rho _{cm} (t)\bigg] |{\bf q}'\rangle\nonumber\\
&&\quad -\lambda\int dr \langle r|\rho _{int} (t)|r\rangle\sum_{i=1}^{N}\sum_{j=1}^{N}
[\Phi({\bf r}_{i}-{\bf r}_{j})-\Phi({\bf r}_{i}-{\bf r}_{j}+{\bf q}-{\bf q}')]
\langle{\bf q}|\rho _{cm} (t)|{\bf q}'\rangle.
\end{eqnarray}

\noindent   Since the nucleons are well-localized, we may write e.g., 
$\int dr \langle r|\rho _{int} (t)|r\rangle\Phi({\bf r}_{i}-{\bf r}_{j})\approx\Phi({\bf z}_{i}-{\bf z}_{j})$ where 
${\bf z}_{i}$ is the mean position of the $i$th nucleon.  Moreover, since the 
nucleii are closely spaced compared to $a=10^{-5}$cm we may, to a good approximation, take them to be continuously distributed and replace   
the double sum in Eq. (A3) by a double integral, obtaining 

\begin{eqnarray}\label{A4} 
&&{\partial\over \partial t}\langle{\bf q}|\rho _{cm} (t)|{\bf q}'\rangle=
-i\langle{\bf q}|\bigg[ {P^{2}\over 2M},\rho _{cm} (t)\bigg] |{\bf q}'\rangle\nonumber\\
&&\qquad -\lambda \bigg( {N\over V}\bigg) ^{2}\int\int_{V} d{\bf z}d{\bf z}' 
[\Phi({\bf z}-{\bf z}')-\Phi({\bf z}-{\bf z}'+{\bf q}-{\bf q}')]
\langle{\bf q}|\rho _{cm} (t)|{\bf q}'\rangle
\end{eqnarray}
	
To see roughly how the collapse part of Eq. (A4) works, suppose that 
$\rho (0)= (1/2)[|\psi_{1}\rangle+|\psi_{2}\rangle][\langle\psi_{1}|+\langle\psi_{2}|]$ and that 
$|\psi_{1}\rangle$, $|\psi_{2}\rangle$ describe two well-separated $(>>a)$ states of a blob so       
$\Phi({\bf z}-{\bf z}'+{\bf q}-{\bf q}')\approx 0$.  Therefore, neglecting the Hamiltonian term, Eq. (A4) 
says that the off-diagonal density elements exponentially decay: 

\[\langle{\bf q}|\rho _{cm} (t)|{\bf q}'\rangle=(1/2)\langle{\bf q}|\psi_{1}\rangle\langle\psi_{2}|{\bf q}'\rangle
e^{-\lambda NN't}.\]

\noindent In this equation, if the dimensions of the blob are $<<a$ 
then $\Phi({\bf z}-{\bf z}')\approx 1$ and $V^{-1}\int d{\bf z}=1$ 
so $N'\approx N$ . If
the dimensions of the blob are $>>a$, $N'\approx$ the number of nucleons in a volume $a^{3}$.  
This collapse rate $\lambda NN'$ is diminished if the blob states overlap 
($\Phi({\bf z}-{\bf z}'+{\bf q}-{\bf q}')\neq 0$).

\subsection{Translational Diffusion Of A Sphere}  

	 We shall apply Eq. (A4) to an ensemble of spheres. Each sphere's 
CM wavefunction (subject to its own sample field $w({\bf x},t)$) reaches an equilibrium size (see Appendix B),  
subject as it is to the Schr\"odinger evolution expansion and 
the collapse interaction contraction, with the center of a new contraction generally located 
off-center from the previous wavefunction center, thereby giving rise to the random walk.     

	We shall use Eq. (A4) to calculate 

\begin{equation}\label{A5}
 \overline {{{\langle {Q^{j}}^{2}\rangle}}}(t)\equiv\int DwP(w)
 {_{w}\langle\psi,t|{Q^{j}}^{2}|\psi,t\rangle _{w}\over _{w}\langle\psi,t|\psi,t\rangle _{w}}
 \equiv Tr [\rho (t) {Q^{j}}^{2}].
\end{equation}	

\noindent In Eq. (A5), $|\psi,t\rangle _{w}$ is the statevector of the sphere at time $t$ 
which evolves under a specific collapse-causing random field $w({\bf x}, t)$, the density matrix is 
$\rho (t)=|\psi,t\rangle _{w}\thinspace _{w}\langle\psi,t|$ and, according to CSL,  
$DwP(w)=Dw_{w}\langle\psi,t|\psi,t\rangle _{w}$ is the probability that the field $w({\bf x}, t)$ appears in nature, 
where $Dw\sim\prod_{{\bf x}, t}dw({\bf x}, t)$ (space-time may be regarded as divided into little cells, in each of which 
$w({\bf x}, t)$ can take on any real value).  In Appendix B we shall calculate 

\begin{equation}\label{A6}
 \overline {{\langle Q^{j}\rangle}^{2}}(t)\equiv\int DwP(w)
 \bigg[{_{w}\langle\psi,t|Q^{j}|\psi,t\rangle _{w}\over _{w}\langle\psi,t|\psi,t\rangle _{w}} \bigg]^{2}
\end{equation}	

\noindent which cannot be expressed as a trace with respect to the density matrix. As shown in Appendix B, the 
spheres we consider are large enough so that the mean 
square packet width $\overline{s^{2}}\equiv\overline{{{\langle [Q^{j}-\langle Q^{j}\rangle]^{2}\rangle}}}$ 
rapidly reaches an equilibrium constant size $<<\overline {{\langle Q^{j}\rangle}^{2}}(t)$.  
Therefore the increase with time $\sim t^{3}$ of $\overline {{\langle Q^{j}\rangle}^{2}}(t)$ 
found here (see Eq. (A10) is solely due to the diffusion of the spheres.  

	To find $ \overline {{{\langle {Q^{j}}^{2}\rangle}}}(t)$ we take successive traces of Eq. (A4):  
 
\begin{mathletters}
\label{A7}
\begin{equation}
{d\over dt} \overline {{{\langle {Q^{j}}^{2}\rangle}}} ={1\over M}\overline{{\langle P^{j} Q^{j}+Q^{j} P^{j}\rangle}}
\end{equation}
\begin{equation}
{d\over dt}{1\over M}\overline{{\langle P^{j} Q^{j}+Q^{j} P^{j}\rangle}} =
{2\over M^{2}}\overline{{\langle {P^{j}}^{2}\rangle}}
\end{equation}
\begin{equation}
{d\over dt}{2\over M^{2}}\overline{{\langle {P^{j}}^{2}\rangle}}={\lambda N^{2}\hbar^{2}f(R/a)\over M^{2}a^{2}}
\end{equation}
\end{mathletters}

\noindent where

\begin{equation}\label{A8}
f(R/a)\equiv{1\over V^{2}}\int\int_{V} d{\bf z}d{\bf z}' \Phi({\bf z}-{\bf z}')
\bigg[1-{(z^{j}-z'^{j})^{2}\over 2a^{2}}\bigg]
\end{equation}

	 Integration of $f$ may be facilitated using Gauss's law to convert the volume integrals to surface integrals: 
	 
\begin{mathletters}\label{A9}
\begin{equation}
f(R/a)={2a^{2}\over V^{2}}\int\int_{V} d{\bf z}d{\bf z}' 
{\bf \nabla}\cdot {\hat {\bf e}}_{j}{\bf \nabla}'\cdot {\hat {\bf e}}'_{j}\Phi({\bf z}-{\bf z}')
=2a^{2}\frac{1}{V^{2}}\int\int_{A} 
d{\bf A}\cdot{\hat {\bf e}}_{j}d{\bf A}'\cdot{\hat {\bf e}}'_{j}\Phi({\bf z}-{\bf z}')
\end{equation}
\begin{equation}
=6\bigg( {a\over R} \bigg)^{4}\bigg[ 1-{2a^{2}\over R^{2}}+\bigg(1+{2a^{2}\over R^{2}}\bigg) e^{-R^{2}/a^{2}}\bigg]
\end{equation}
\begin{equation}
\thinspace_{\overrightarrow {R<<a}}1, \qquad f(1)=.62, \qquad_{\overrightarrow {R>>a}}6\bigg({a\over R}\bigg)^{4}.  
\end{equation}
\end{mathletters}
	 
\noindent $f$ is a monotonically decreasing function of its argument.  

	It follows from Eqs. (A.7) that  
	
\begin{equation}\label{A10}
 \overline {{{\langle {Q^{j}}^{2}\rangle}}} =
 \langle \bigg( Q^{j}+\frac{P^{j}t}{M}\bigg)^{2}\rangle (0)+{\lambda\hbar^{2}f(R/a)t^{3}\over 6 m^{2}a^{2}} 
\end{equation}	
	
\noindent which is the result quoted in Eq. (3.4). The diffusion term in Eq. (A10) can be understood as follows.  
$d^{3}\overline {{{\langle {Q^{j}}^{2}\rangle}}}(t)/dt^{3}$ is proportional to the square 
of the collapse-induced velocity $(\hbar/ Ma)^{2}$ multiplied by the effective collapse rate. For 
$R<<a$, the collapse rate is $\sim\lambda N^{2}$, giving rise to Eq. (A10) with $f=1$. For 
$R>>a$, as we have previously shown\cite{GPR}, the collapse rate 
is $\sim\lambda\cdot$(number of particles in a volume $a^{3})\cdot($number of uncovered particles).  
That is, we imagine the sphere in a superposition of two states displaced from each other by a certain distance so 
the two images of the sphere overlap: the ``uncovered" particles are those in the region of no overlap.  In this 
case we suppose the displacement distance is $a$.  Then, the (number of uncovered particles)$\approx (N/V)a$(the 
surface area $A$ of the sphere).  Thus we find the expression 

\[\sim (\hbar/ Ma)^{2} \lambda (Na^{3}/V)(NAa/V)\sim\lambda(\hbar/ ma)^{2}(a/R)^{4}\sim\lambda(\hbar/ ma)^{2}f.\]

\subsection{Translational Diffusion Of A Disc} 

	In the case of a disc undergoing translational diffusion, 
$f$ depends upon its orientation.  If the disc is of radius $L$ and thickness $b$, 
for motion perpendicular to the disc face it follows from Eq. (A9a) (which is applicable 
to an arbitrarily shaped object) that 

\begin{equation}\label{A11}
f=4\bigg(\frac{2a}{L}\bigg)^{4}\bigg(\frac{2a}{b}\bigg)^{2}\bigg[1-e^{-b^{2}/4a^{2}}\bigg]
\int_{0}^{L/2a}xdx\int_{0}^{L/2a}x'dx'e^{-(x^{2}+x'^{2})}I_{0}(2xx')
 \end {equation} 

\noindent For example, for $(b/2a)^{2}<<1$, $f\approx 1$ for $(L/2a)^{2}<<1$ and 
$f\rightarrow (2a/L)^{2}$ for $(L/2a)^{2}>>1$. 
For motion parallel to the disc edge, 

\begin{equation}\label{A12}
f=\bigg(\frac{2a}{L}\bigg)^{2}e^{-L^{2}/2a^{2}}I_{1}(L^{2}/2a^{2})\bigg(\frac{2a}{b}\bigg)^{2}
\bigg[\frac{b}{2a}\int_{-b/2a}^{b/2a}dxe^{-x^{2}}-1+e^{-(b/2a)^{2}}\bigg]
 \end {equation} 

\noindent where $f\rightarrow(4/\sqrt{\pi})(a/L)^{3}$ for $(b/2a)^{2}<<1$ and $(L/2a)^{2}>>1$.
  
	Eqs. (A11, A12) can also be understood as proportional to the effective collapse rate.   
We use an alternative expression for the rate, equivalent to that given in the previous paragraph, 
 appropriate for an object with a dimension ($b$ in this case) less than $a$.  It is  
 rate $\sim$ (number of particles in a cell)$^{2}\cdot$(number of uncovered cells), where a cell 
 is a cube of dimension $a$ on each side. Here each cell has occupied  
 volume $ba^{2}$ so the number of particles/cell $=(Nba^{2}/\pi L^{2}b)\sim (a/L)^{2}$.  
For displacement $a>>b$ perpendicular to the face, all the cells---$\pi L^{2}/a^{2}$ of them---are uncovered, 
so we obtain the collapse rate $\sim (a/L)^{4}(L/a)^{2}=(a/L)^{2}$. For motion parallel to the face,
displacement $a$ uncovers the cells lying on the circumference of the disc $\approx 2\pi L/a$ of them, giving 
the collapse rate $\sim (a/L)^{4}(L/a)=(a/L)^{3}$.

\section {Wavepacket Width Of Center Of Mass in CSL}

	The CSL evolution equation for the normalized statevector in Stratonovitch form (so manipulations 
can be performed using the usual rules of calculus) is 

\begin{mathletters}
\begin{equation}\label{B1a}
{d\over dt}|\psi,t\rangle_w=\bigg\{ -iH+
\bigg[\int d{\bf x} G({\bf x})w({\bf x},t)-
\lambda \int d{\bf x} \big[ G^{2}({\bf x})-
\thinspace _{w}\langle \psi,t|G^{2}({\bf x})|\psi,t\rangle_w\big]\bigg]\bigg\}|\psi,t\rangle_w
\end{equation}
\begin{equation}\label{B1b}
G({\bf x})\equiv{1\over (\pi a^{2})^{3/4}}\sum_{j=1}^{N} \bigg[ e^{-({\bf X}_{j}-{\bf x})^{2}/2a^{2}}\thinspace 
-\thinspace_{w}\langle \psi,t|e^{-({\bf X}_{j}-{\bf x})^{2})/2a^{2}}|\psi,t\rangle_w\bigg]
\end{equation}
\end{mathletters}

\noindent where ${\bf X}_{j}$ is the position operator for the jth nucleon, 
 $w({\bf x},t)=dB({\bf x},t)/dt$ is standard white noise and $B({\bf x},t)$ is standard Brownian motion 
($\overline {w({\bf x},t)} =0$, $\overline {w({\bf x},t)w({\bf x}',t')}=
\lambda \delta ({\bf x}-{\bf x}')\delta (t-t'))$.  We extract the equation for the CM wavefunction 
just as in Appendix A whose notation is 
used here (Eq. (A1) for the density matrix can readily be derived from Eq. (B1)).  
Again, we assume that the statevector can be written as a direct product of the 
internal statevector $|\psi_{int},t\rangle$ and the CM statevector $|\phi,t\rangle_w$. We suppose that 
$|\psi_{int},t\rangle$ obeys the usual Schrodinger equation (thereby neglecting the CSL excitation of atoms 
and nucleii) with Hamiltonian $H_{int}$, so the complete Hamiltonian is $H={\bf P}^{2}/2M+H_{int}$, 
where ${\bf P}$ is the CM momentum operator. Using this in Eq. (B.1) 
with ${\bf X}_{j}={\bf R}_{j}+{\bf Q}$, multiplying by $\int dr\langle\psi_{int},t|r\rangle\langle r|$, 
employing the localized nature of nucleons so, e.g., 
$\int dr|\langle\psi_{int},t|r\rangle|^{2}F({\bf r}_{j})\approx F({\bf z}_{j})$ where ${\bf z}_{j}$ 
is the mean position of the $j$th nucleon, and then approximating 
$\sum_{j} F({\bf z}_{j})\approx (N/V)\int_{V}d{\bf z}F({\bf z})$ results in

 \begin{mathletters}
\begin{equation}\label{B2a}
{d\over dt}\langle {\bf q}|\phi,t\rangle_w=\bigg\{ -i{\bigtriangledown ^{2}\over 2M}+
\bigg[ \int d{\bf x} g({\bf x}-{\bf q})w({\bf x},t)-
\lambda \int d{\bf x} \
\big[ g^{2}({\bf x}-{\bf q})-
\thinspace _{w}\langle \psi,t|g^{2}({\bf x}-{\bf q})|\psi,t\rangle_w\big]\bigg]\bigg\}|\phi,t\rangle_w
\end{equation}
\begin{equation}\label{B2b}
g({\bf x}-{\bf q})\equiv {1\over (\pi a^{2})^{3/4}}\bigg( {N\over V} \bigg) \int_{V} d {\bf z}
\bigg[ e^{-({\bf z}+{\bf q}-{\bf x})^{2}/2a^{2}}\thinspace 
-\thinspace_{w}\langle \phi,t|e^{-({\bf z}+{\bf q}-{\bf x})^{2}/2a^{2}}|\phi,t\rangle_w\bigg]
\end{equation}
\end{mathletters} 

\noindent (Eq. (A4) for the CM density matrix can readily be derived from Eq. (B2)).

\subsection{A Sphere's Equilibrium CM Wavepacket Width And The Time To Reach It}

	Eq. (B2), applied to a sphere of radius $R$, is the starting point for our calculation.   
We shall consider only cases where, for
each $|\phi,t\rangle_w$, the squared wavepacket width $s^{2}(t)\equiv 
\thinspace_{w}\langle \phi,t|{Q^{j}}^{2}|\phi,t\rangle_w-{\thinspace_{w}\langle \phi,t|Q^{j}|\phi,t\rangle_w}^{2}
=\thinspace_{w}\langle \phi,t|[{Q^{j}-\langle Q^{j}\rangle}]^{2}|\phi,t\rangle_w$ is much less than 
$a^{2}$ (note that $s$ has no subscript $j$ because we assume its initial spherical symmetry 
which is maintained thereafter). At the end of section IIIit is shown   
that $s<<a$ implies $R>10^{-6}$cm.
 
	We expand the exponents in Eq. (B2b) in 
powers of $[Q^{j}-\langle Q^{j}\rangle]/a$, retaining only the leading term:

\begin{equation}\label{B3}
g({\bf x}-{\bf q})\approx {1\over (\pi a^{2})^{3/4}}\bigg( {N\over V} \bigg) \int_{V} d {\bf z}
e^{-({\bf z}+\langle{\bf Q}\rangle -{\bf x})^{2}/2a^{2})}
({\bf z}+\langle{\bf Q}\rangle -{\bf x})\cdot({\bf q}-\langle{\bf Q}\rangle)/a^{2}.
\end{equation}
	 
	The solution of Eq. (B2a), for a short time $\Delta t$, can be written (with use of Eq. (B3)) as 
	
\begin{eqnarray}\label{B4}
&& \langle {\bf q}|\phi,t+\Delta t\rangle_{w}
=\exp\bigg[ i\Delta t {\bigtriangledown ^{2}\over 2M}
-{1\over a^{2}(\pi a^{2})^{3/4}}({\bf q}-\langle{\bf Q}\rangle)\cdot
  \bigg( {N\over V} \bigg)\int_{V} d {\bf z}\int d {\bf x}({\bf z}-{\bf x})e^{-({\bf z}-
{\bf x})^{2}/2a^{2}}dB({\bf x},t) \nonumber\\
&&\qquad\qquad\qquad\qquad\qquad\qquad-\lambda \Delta t {N^{2}\over 2a^{2}}f(R/a) 
\big[({\bf q}-\langle{\bf Q}\rangle)^{2}-
\thinspace _{w}\langle \psi,t|({\bf q}-\langle{\bf Q}\rangle)^{2}|\psi,t\rangle_w\big]\bigg]\langle 
{\bf q}|\phi,t\rangle_{w}
\end{eqnarray}

\noindent (note the replacement of ${\bf x}-\langle{\bf Q}\rangle$ by ${\bf x}$ as dummy integration variable and the 
concommitant use of translation invariance of $w({\bf x},t))$§ where $f(R/a)$ is given by Eq. (A8).  Eq. (B4)  
shows that a gaussian wavefunction at time $t$ is taken into a gaussian wavefunction at time $t+\Delta t$.

	Although we could deal with a more general class of wavefunction, the results are the same and the argument is simpler  
if we restrict ourselves to the complex gaussian wavefunction   

\begin{equation}\label{B5}
\langle {\bf q}|\phi,t\rangle_{w}=A e^{-({\bf q}-{\bf b})^{2}/4\sigma^{2}}, 
\qquad A\equiv (2\pi \sigma^{2}{\sigma ^ {*}}^{2}/\sigma_{R}^{2})^{-3/4}e^{-b_{I}^{2}/4\sigma_{R}^{2}}. 
\end{equation}

\noindent In Eq. (B5), ${\bf b}={\bf b}_{R}+i{\bf b}_{I}$, $\sigma^{2}=\sigma_{R}^{2}+i\sigma_{I}^{2}$ 
are complex functions of time.  Using this wavefunction one can calculate various 
expectation values involving the CM position and momentum. It follows from  Eq. (B5) that 

\begin{eqnarray}\label{B6}
&&\qquad\qquad\qquad\qquad\qquad\qquad
|\langle {\bf q}|\phi,t\rangle_{w}|^{2}=(2\pi s^{2})^{-3/4} e^{-({\bf q}-\langle{\bf Q}\rangle)^{2}/2 s^{2}},\nonumber\\ 
&&\quad \langle{\bf Q}\rangle ={\bf b}_{R}+{\bf b}_{I}\sigma_{I}^{2}/\sigma_{R}^{2}, 
\quad \langle{\bf P}\rangle ={\bf b}_{I}2\sigma_{R}^{2},
\quad s^{2}\equiv\langle (Q^{j}-\langle Q^{j}\rangle)^{2}\rangle=\sigma_{R}^{2}+\sigma_{I}^{4}/\sigma_{R}^{2}, 
\quad\langle (P^{j}-\langle P^{j}\rangle)^{2}\rangle=1/4 \sigma_{R}^{2}.
\end{eqnarray}

\noindent	We note that 

\[ e^{i\Delta t \bigtriangledown ^{2}/2M}\langle {\bf q}|\phi,t\rangle_{w}=
A e^{-({\bf q}-{\bf b})^{2}/4[\sigma^{2}+i(\Delta t/2M)]}
\approx e^{({\bf q}-{\bf b})^{2}i(\Delta t/8M\sigma^{4})}\langle {\bf q}|\phi,t\rangle_{w}. \]

\noindent Putting this into Eq. (B4) and equating the coefficients of ${\bf q}^{2}$ and ${\bf q}$ results in 

 \begin{mathletters}
\begin{equation}\label{B7a}
{1\over 4}{d\over dt}{1\over\sigma^{2}}=-{i\over 8M\sigma^{4}}+ {\lambda N^{2}\over2a^{2}}f
\end{equation}
\begin{equation}\label{B7b}
{1\over 2}{d\over dt}{{\bf b}\over\sigma^{2}}=-{i{\bf b}\over 4M\sigma^{4}}+ 
{\lambda N^{2}\over a^{2}}\langle {\bf Q}\rangle f-
{1\over a^{2}(\pi a^{2})^{3/4}}
{N\over V}\int_{V} d {\bf z}\int d {\bf x}({\bf z}-{\bf x})e^{-({\bf z}-{\bf x})^{2}/2a^{2}}w({\bf x},t)
\end{equation}
\end{mathletters} 

	First, consider Eq. (B7a): 
	
\begin{equation}\label{B8}
{d\over dt}\sigma^{2}=-{i\over 2M}- {2\lambda N^{2}\over a^{2}}f\sigma^{4}
\end{equation}

\noindent It has no stochastic part and may be immediately solved.  
As $t\rightarrow \infty$, where $d\sigma^{2}(t)/dt=0$, 
according to Eq. (B8),

	\[ \sigma^{2}(\infty)=(a/2N)(\hbar/2M\lambda f)^{1/2}(1+i) \]

\noindent so, by Eq. (B6), the asymptotic squared wavepacket width is 

\begin{equation}\label{B9}
s^{2}(\infty)\equiv s_{\infty}^{2} =(a/N)(\hbar/2M\lambda f)^{1/2}.
\end{equation}
 
\noindent This result can be obtained by a simple physical argument given in section III (following Eq. (3.5)).

	Introducing $\sigma^{2}(\infty)$ into Eq. (B8), together with 

\begin{equation}\label{B10}
\tau_{s}\equiv Ms_{\infty}^{2}/\hbar
\end{equation}

\noindent converts Eq. (B8) to 

\begin{equation}\label{B11}
{d\over d(t/\tau_{s})}\bigg({\sigma\over s_{\infty}}\bigg) ^{2}=-{i\over 2}- \bigg({\sigma\over s_{\infty}}\bigg) ^{4}
\end{equation}

\noindent with solution

\begin{equation}\label{B12}
\sigma^{2}(t)=s_{\infty}^{2}{(1+i)\over 2}
\Bigg[{{\sigma^{2}(0)\over s_{\infty}^{2}}\big[ e^{t(1+i)/\tau_s}+1\big]+{(1+i)\over 2}\big[ e^{t(1+i)/\tau_s}-1\big]
\over {\sigma^{2}(0)\over s_{\infty}^{2}}\big[ e^{t(1+i)/\tau_s}-1\big]+{(1+i)\over 2}\big[ e^{t(1+i)/\tau_s}+1\big]}\Bigg]
\end{equation}

\noindent which shows the approach to equilibrium.  

	Thus we have achieved the main purpose of this appendix, to obtain the cm wavepacket equilibrium width (B9) 
and the characteristic time to reach that width (B10). We emphasize that these results apply to {\it every} 
cm wavepacket, since they are independent of the particular realization of the fluctuating field $w$ 
encountered by a sphere.  

\subsection{CM Translational Diffusion Revisited}

However, we do have Eq. (B7b) which does depend upon $w$ and 
which gives us, via Eq. (B6), each individual cm wavefunction's mean position $\langle{\bf Q}\rangle$ 
(and mean momentum $\langle{\bf P}\rangle$), enabling us to understand in detail the ensemble average 
$\overline {{{\langle {Q^{j}}^{2}\rangle}}}$ in Eq. (A10).  

	We first note that the stochastic term in Eq. (B7b) has no dependence on the dynamical variables 
$\sigma^{2}$ and ${\bf b}$.  Since the ensemble average of the product of this term's $i$th and $j$th components is 
$\delta_{ij}\lambda N^{2}f\delta (t-t')/2a^{2}$ we may write the term as 
$(\lambda N^{2}f/2a^{2})^{1/2}{\bf w}(t)$ where the $w_{j}(t)$'s are independent white noise, 
$\overline{w_{i}(t)w_{j}(t')}=\delta_{ij}\delta (t-t')$.  We may then write Eq. (B7b) as

\begin{equation}\label{B13}
{d\over dt}{\bf b}=-{i\sigma^{4}\over \sigma_{R}^{2}}{{\bf b}_{I}\over s_{\infty}^{2}\tau_{s}} 
+{\sigma^{2}{\bf w}(t)\over s_{\infty}\tau_{s}^{1/2}}.
\end{equation}

	Suppose we follow an individual wavefunction for sufficient time $>>\tau_{s}$ until (say, at time $t=0$) it achieves 
its equilibrium width $s_{\infty}$ with $\sigma^{2}=s_{\infty}^{2}(1+i)/2$ so Eq. (B13) simplifies to 

\begin{equation}\label{B14}
d{\bf b}={{\bf b}_{I}\over \tau_{s}}dt 
+{(1+i)\over 2}{s_{\infty}\over\tau_{s}^{1/2}}d{\bf B}(t)
\end{equation}

\noindent where $B(t)$ is Brownian motion ({\bf w}(t)={d\bf B}(t)/dt). The solution of Eq. (B.14) is 

\begin{equation}\label{B15}
b^{j}_{R}(t)={s_{\infty}\over 2\tau_{s}^{3/2}}\int_{0}^{t}dt'B^{j}(t')+{s_{\infty}\over 2\tau_{s}^{1/2}}B^{j}(t), \quad
b^{j}_{I}(t)={s_{\infty}\over 2\tau_{s}^{1/2}}B^{j}(t)
\end{equation}

\noindent (we have assumed $b^{j}_{R}(0)=b^{j}_{I}(0)=0$). 

	It follows from Eqs. (B15) and (B6) that 
	
\begin{equation}\label{B16}	
\langle{\bf Q}\rangle=\frac{s_{\infty}}{2\tau_{s}^{3/2}}\int_{0}^{t}dt'{\bf B}(t')+{s_{\infty}\over \tau_{s}^{1/2}}{\bf B}(t),
\qquad \langle{\bf P}\rangle=\frac{1}{2s_{\infty}\tau_{s}^{1/2}}{\bf B}(t)	
\end{equation}
	
\noindent which explicitly shows the diffusive nature of $\langle{\bf Q}\rangle$ and $\langle{\bf P}\rangle$.

	We can now find
$\overline{\langle {Q^{j}}^{2}\rangle}=s_{\infty}^{2}+\overline{\langle Q^{j}\rangle^{2}}$ and compare with Eq. (A10).   
 Recalling that $\overline{{B^{j}}^{2}(t)}=t$ and $\overline{B^{j}(t)B^{j}(t')}=$min$(t,t')$, we obtain 

\begin{equation}\label{B17}
\overline{\langle {Q^{j}}^{2}\rangle}=s_{\infty}^{2}+s_{\infty}^{2}\bigg[{t\over \tau_{s}}+
{t^{2}\over 2\tau_{s}^{2}}+{t^{3}\over 12\tau_{s}^{3}}\bigg]
\end{equation}

\noindent which is identical to  Eq. (A10) for a wavefunction which has equilibrium width $s_{\infty}$ at $t=0$.  

\section{Rotational Diffusion in CSL}

	Starting with Eq. (A1) for the evolution of the density matrix in CSL, we follow the lines of 
argument in Appendix A. We assume here that the cm of a blob of matter is fixed but that it is 
free to rotate about a fixed axis through an angle represented by the operator $\Theta$ with angular 
momentum operator ${\cal L}$. We also assume that the density matrix is the direct product of 
the internal density matrix and the orientation density matrix $\rho_{ang}$.  We obtain, analogous to Eq. (A4), 
 
\begin{eqnarray}\label{C1} 
&&{\partial\over \partial t}\langle\theta|\rho _{ang} (t)|\theta'\rangle =
-i\langle\theta|\bigg[ {{\cal L}^{2}\over 2I},\rho_{ang}(t)\bigg] |\theta'\rangle\nonumber\\
&&\qquad -\lambda \bigg( {N\over V}\bigg)^{2}\int\int_{V} d{\bf z}d{\bf z}' 
[\Phi({\bf z}(0)-{\bf z}'(0))-\Phi({\bf z}(\theta)-{\bf z}'(\theta'))]
\langle\theta|\rho_{ANG}(t)|\theta'\rangle
\end{eqnarray}
	
\noindent where, denoting the rotation axis by $z_{3}$, 
 
\begin{equation}\label{C2}
\Phi({\bf z}(\theta)-{\bf z}'(\theta'))=\exp -{1\over 4a^{2}}[{\bf z}^{2}+{\bf z}'^{2}
-2(z_{1}z'_{1}+z_{2}z'_{2})\cos (\theta-\theta ')-2(z_{1}z'_{2}-z_{2}z'_{1})\sin (\theta-\theta ')-2z_{3}z'_{3}].	
\end{equation}

	To find $\overline{\langle\Theta^{2}\rangle}(t)$, in analogy to Eqs. (A7), we take successive 
traces of Eq. (C1):
 
\begin{mathletters}
\label{all C3}
\begin{equation}
{d\over dt} \overline {{{\langle {\Theta}^{2}\rangle}}} ={1\over I}\overline{{\langle {\cal L} \Theta +\Theta {\cal L}\rangle}}
\end{equation}
\begin{equation}
{d\over dt}{1\over I}\overline{{\langle {\cal L} \Theta + \Theta {\cal L}\rangle}} =
{2\over I^{2}}\overline{{\langle {{\cal L}}^{2}\rangle}}
\end{equation}
\begin{equation}
{d\over dt}{2\over I^{2}}\overline{{\langle {{\cal L}}^{2}\rangle}}=
{\lambda\over 2} \bigg[{\hbar\over ma^{2}}\bigg]^{2}f_{ROT}
\end{equation}
\end{mathletters}

\noindent where ${\bf z}_{\bot}\equiv(z_{1},z_{2})$ and $f_{ROT}$ is the dimensionless geometrical factor 

\begin{equation}\label{C4}
f_{ROT}=2\bigg[{\frac{Ma}{IV}}\bigg]^{2}\int\int_{V} d{\bf z}d{\bf z}'
\big[{\bf z}_{\bot}\cdot{\bf z}'_{\bot}-{1\over 2a^{2}}({\bf z}_{\bot}\times{\bf z}'_{\bot})^{2}\big]\Phi({\bf z}-{\bf z}'). 
\end{equation}

	To see that (C3) vanishes if the blob is rotationally symmetric about 
the $z$-axis (i.e., a sphere or a disc with the $z$-axis perpendicular to its face), we write (C4) as

\begin{equation}\label{C5}
 f_{ROT}=-4\bigg[{\frac{Ma^{2}}{IV}}\bigg]^{2}
 \int_{V}d{\bf z}({\bf z}_{\bot}\times\nabla_{{\bf z}_{\bot}})^{2}\int_{V} d{\bf z}'\Phi({\bf z}-{\bf z}').
\end{equation}

\noindent Rotational symmetry implies that the integral over ${\bf z}'$ is just a function of ${\bf z}^{2}$ 
and then the integral vanishes since $({\bf z}_{\bot}\times\nabla_{{\bf z}_{\bot}}){\bf z}^{2}=0$.

	It follows from Eqs. (C3) that the mean square angular diffusion has the time dependence 
	
\begin{equation}\label{C6}	
\overline {{{\langle {\Theta}^{2}\rangle}}}=\langle\bigg( \Theta+\frac{{\cal L}t}{I}\bigg) ^{2}\rangle(0)+
\lambda \frac{t^{3}}{12}\bigg[\frac{\hbar}{ma^{2}}\bigg]^{2}f_{ROT}.	
\end{equation}	

	We wish apply Eq. (C6) to a disc of radius $L$ and thickness $b$ ($I=(ML^{2}/4)[1+(b^{2}/3L^{2}]$), 
with the rotation axis parallel to the face of the disc, so $f_{ROT}$ must be calculated for this case. 
The double volume integral in Eq. (C5) can be converted to a double integral over the surface of the disc 
by using the divergence theorem:

\begin{equation}\label{C7}
 f_{ROT}(\alpha, \beta)=\bigg[\frac{2}{[1+(\beta^{2}/3\alpha^{2})]\beta\alpha^{4}}\bigg]^{2}
 \int_{A}d{\bf A}\cdot ({\bf r}\times {\bf k})\int_{A}d{\bf A}'\cdot ({\bf r}'\times {\bf k})e^{-({\bf r}-{\bf r}')^{2}}
 \end{equation}

\noindent where ${\bf k}$ is the unit vector along the axis of rotation and $\alpha\equiv L/2a$, $\beta\equiv b/2a$. 
Calling the contribution of the two disc faces $f_{1}$, the two disc edges $f_{2}$, and 
the edge-face contribution $f_{3}$, we obtain

\begin{mathletters}
\label{all C8}
\begin{equation}
f_{ROT}(\alpha, \beta)=\bigg[\frac{4}{[1+(\beta^{2}/3\alpha^{2})]\beta\alpha^{4}}\bigg]^{2}[f_{1}+f_{2}+f_{3}]
\end{equation}
\begin{equation}
f_{1}=[1-e^{-\beta^{2}}]\int_{0}^{\alpha}r^{2}dr\int_{0}^{\alpha}r'^{2}dr'I_{1}(2rr')e^{-(r^{2}+r'^{2})}
\end{equation}
\begin{equation}
f_{2}=(1/2)\alpha^{2}e^{-2\alpha^{2}}I_{1}(2\alpha^{2})\int_{-\beta/2}^{\beta/2}ydy\int_{-\beta/2}^{\beta/2}y'dy'e^{-(y-y')^{2}}
\end{equation}
\begin{equation}
f_{3}=-2\alpha e^{-\alpha^{2}}\int_{0}^{\alpha}r^{2}dre^{-r^{2}}I_{1}(2\alpha r)\int_{-\beta/2}^{\beta/2}ydye^{-(.5\beta-y)^{2}}
\end{equation}
\end{mathletters}

\noindent where $I_{1}$ is the Bessel function. 

	A graph of $f_{ROT}(\alpha, \beta)$ vs $\alpha$ parametrized by 
various values of $\beta$ is given in FIG. 1. 
We note that, for a thin disc ($\beta<<\alpha$) with $\beta<<1$, $f_{1}$ is the leading term in Eq. (C8a)  which becomes 

\begin{equation}\label{C9}	
f_{ROT}(\alpha)\approx \bigg(\frac{2}{\alpha}\bigg)^{4}
\int_{0}^{\alpha}r^{2}dr\int_{0}^{\alpha}r'^{2}dr'I_{1}(2rr')e^{-(r^{2}+r'^{2})}.
\end{equation}

\section{Thermal Radiation Viscosity Factor For A Dielectric Sphere}

	In order to compare the Brownian diffusion of an object in 
a thermal radiation bath with CSL diffusion, it is only necessary to 
find the viscosity factor $\xi$ for this situation and put it into the Brownian motion equations of section II.  
An object moving with respect to thermal radiation with speed $v$ 
feels a drag force $-\xi v$ because it receives 
more momentum from the photons it approaches than from those from which it recedes.  We have not 
been able to find $\xi$ for a dielectric sphere in the literature (the closest has been the force on an oscillator\cite{Hoye}) so 
we give it here. Actually, after this Appendix was written, we decided that the experiment we propose would concern 
a conducting disc rather than a dielectric sphere! The  result for a conducting sphere is not quite the same as that for 
a dielectric sphere with dielectric constant equal to infinity: although that is the appropriate limit for electric field 
behavior, a conducting sphere's magnetic behavior is also 
important in considering the scattering cross-section of electromagnetic radiation (necessary for this calculation).  
And, of course, a sphere is not a disc. However, the result obtained for $\xi$ will be representative, i.e., 
the same up to a numerical factor not too far from 1, when the dielectric constant goes to infinity and the radius of the 
sphere is replaced by the radius of the disc. 

\subsection{Viscosity Factor For a Mirror}

	For expositional ease and purposes of comparison we shall first obtain Einstein's result for a mirror
moving perpendicular to its face\cite{EinsteinandHopf,Einstein}, as seen from the laboratory frame 
in which the radiation is thermal.  We shall use properties of photons (which Einstein had not yet obtained 
as he was in the process of establishing these, so he used classical electromagnetism).

	As is well known, at temperature $T$ the mean number of photons in a mode of frequency $\nu_{0}$  (the subscript 0 refers to the laboratory frame) 
is $[\exp\beta h \nu_{0}-1]^{-1}$ (where $\beta\equiv (kT)^{-1}$).  Since the number of photon modes/vol 
of frequency $\nu_{0}$ in the range $d\nu_{0}$ moving in a direction ($\theta_{0}$, $\phi_{0}$) 
within solid angle $d\Omega_{0}$ is $2\nu_{0}^{2}d\nu_{0}d\Omega_{0}/c^{3}$ (the factor 2 is for the two polarizations), 
the mean photon number/vol-freq-solid angle is

 \begin{equation}\label{D1}
	n(\nu_{0})=2(\nu_{0}^{2}/c^{3})[\exp\beta h \nu_{0}-1]^{-1}
\end{equation}

	First we find the momentum transferred to the mirror by a colliding photon. 
Einstein considered a mirror which, in its rest frame, is perfectly reflecting only for frequencies in the range 
$(\nu, \nu+d\nu)$ (no subscript refers to the rest frame of the mirror) and perfectly transmitting otherwise.  
Let the mirror (of area $A$) move 
in the $z$-direction with speed $v$ away from a photon of momentum $p_{0}=h\nu_{0}/c$ whose 
direction of motion makes an angle $\theta_{0}$ with respect to the $z$-axis.  
The photon will be reflected only if it has frequency $\nu$ in the rest frame of the mirror.  
From the energy and momentum transformations of special relativity (all calculations are to order $v/c$), 

\begin{equation}\label{D2}
	\nu=\nu_{0}[1-(v/c)\cos\theta_{0}], \qquad \nu\cos\theta =\nu_{0}[\cos\theta_{0}-(v/c)]
\end{equation}

\noindent where $\theta$ is the angle the photon makes with the normal to the mirror in the mirror's rest frame.
In this frame the photon's incident and outgoing (normal) momenta are respectively 
$(h\nu /c)\cos\theta$ and $-(h\nu /c)\cos\theta$ so the momentum imparted (normal) to the mirror is 
$\Delta P=2(h\nu /c)\cos\theta$.  The difference of momenta of a nonrelativistic object is a Galilean invariant.  
Therefore, $\Delta P=\Delta P_{0}$ which, by (D2) may be written as 

\begin{equation}\label{D3}
\Delta P_{0}=2(h\nu_{0}/c)[\cos\theta_{0}-(v/c)].	
\end{equation}

	Next we find the number of these photons colliding with the mirror in time $dt$.
In the mirror rest frame this is  ${\bf J}\cdot {\bf A}dt$ where ${\bf J}$ is the particle number flux and ${\bf A}=A{\bf {\hat z}}$ 
with $A$ the area of the mirror.  This is
the same number which collides with the mirror in the laboratory frame in time $dt$.  
The four-current transformation equation gives 

\begin{equation}\label{D4} 
{\bf J}\cdot {\bf A}=({\bf J}_{0}-\rho_{0}{\bf v})\cdot {\bf A}=n(\nu_{0})d\nu_{0}d\Omega_{0}(c\cos\theta_{0}-v)A.
\end{equation}

 Thus, by Eqs. (D3, D4), 
the momentum transferred to the mirror in the laboratory frame in time $dt$, expressed in 
laboratory frame coordinates, is 

\begin{equation}\label{D5}
-vdtd\xi\equiv{\bf J}\cdot {\bf A}dt\Delta P_{0}=
n(\nu_{0})d\nu_{0}d\Omega_{0}c[\cos\theta_{0}-(v/c)]Adt2(h\nu_{0}/c)[\cos\theta_{0}-(v/c)].
\end{equation}

	It remains to integrate Eq. (D5) over all $\Omega_{0}$ but, first, we must express $\nu_{0}$ in terms of $\nu$ and 
$\theta_{0}$.  From the inverse of Eq. (D2) 
we have $\nu_{0}=\nu[1+(v/c)\cos\theta_{0}]$ so we obtain 
 
\[
d\nu_{0}=d\nu[1+(v/c)\cos\theta_{0}], \qquad \nu_{0}n(\nu_{0})=
\nu[1+\frac{v}{c}cos\theta_{0}]\eta\bigg(\nu[1+\frac{v}{c}cos\theta_{0}]\bigg)=\nu n(\nu)+(\nu n(\nu))'\nu(v/c)\cos\theta_{0}+o(v/c)^{2}.  
\]

\noindent Then, we must remember that the above analysis is predicated upon the mirror receding 
from these photons (so the range of $\theta_{0}$ is (0, $\pi /2$)). The momentum 
imparted by the photons on the other side of the mirror is given by the negative of the right hand side of 
Eq. (D5) with the replacement 
$v\rightarrow-v$.  Thus, the contribution from all photons to the force is 

\begin{eqnarray}\label{D6} 
-vd\xi &&=d\nu 2hA\int_{0}^{\pi/2}d\Omega_{0}\bigg\{ 
\nu n(\nu)[\cos^{2}\theta_{0}-(v/c)(2\cos\theta_{0}-\cos^{3}\theta_{0})]+
(\nu n(\nu))'\nu(v/c)\cos^{3}\theta_{0}\bigg\}\nonumber\\
&&\qquad-(v\rightarrow-v)\nonumber\\
&&=-vd\nu2\pi(h/c)A[3\nu n(\nu)-\nu(\nu n(\nu))'].
\end{eqnarray}

	This is Einstein's result.  Putting Eq. (D1) for $n(\nu)$ ($\nu=\nu_{0}$ to zeroth order in $v/c$) 
into Eq. (D6) yields 
	
\begin{equation}\label{D7}
d\xi=4\pi \bigg(\frac{\nu}{c}\bigg)^{3}\bigg(\frac{h\nu}{kT}\bigg)\bigg(\frac{h}{c}\bigg)
\frac{e^{\beta h\nu}}{[e^{\beta h\nu}-1]^{2}}Ad\nu.  
\end{equation}

\noindent Of course, $\nu$ may be integrated over to obtain the viscosity factor for a mirror which 
is a perfect reflector at all frequencies:

\begin{equation}\label{D8}
\xi=4\pi h\bigg(\frac{kT}{hc}\bigg)^{4}A\int_{0}^{\infty}dz\frac{z^{4}e^{z}}{(e^{z}-1)^{2}}=
\frac{2\pi^{2}}{15}\hbar\bigg(\frac{kT}{\hbar c}\bigg)^{4}A.     
\end{equation}

\subsection{Viscosity Factor For A Dielectric Sphere}

	Our discussion for a dielectric sphere (dielectric constant $\epsilon$, radius $R$,  
moving in the $z$-direction with speed $v$) exactly parallels that for the mirror. 

First we find the momentum transferred to the sphere by a colliding photon. In the rest frame of the sphere, 
the scattered radiation has a dipole pattern so the radiation scattered in two opposite directions carries no net momentum.  Thus, 
for radiation of frequency $\nu$, insofar as momentum transfer is concerned, the sphere acts 
like an absorber (of area equal to the total scattering cross-section $\sigma(\nu)$). Thus, the momentum {\it effectively} imparted (i.e., on average) 
in the $z$-direction by an incident colliding photon 
is $\Delta P=(h\nu/c)\cos\theta$.  As in our previous discussion, since $\Delta P=\Delta P_{0}$, 
the effective momentum imparted by a single photon in the laboratory frame is 1/2 of the value given in Eq. (D3).  

	Next we find the number of these photons colliding with the sphere in time $dt$.  In the 
sphere rest frame this is $J\sigma (\nu )dt$, where $J$ is the number flux: this is the same 
number that collides with the sphere in the laboratory frame in time $dt$. To express this number in terms of laboratory frame
variables, we note that $J/c$ is the zeroth component of the current 4-vector and 
$J\cos\theta$ is the component along the $z$-axis.  Therefore the Lorentz transformation of the 
zeroth component of the current 4-vector is $J/c=[J_{0}/c-(v/c^{2})J_{0}\cos\theta_{0}]$ or, 
substituting for $J_{0}$,  

\begin{equation}\label{D9}
 J\sigma(\nu)dt=n(\nu_{0})d\nu_{0}d\Omega_{0}c[1-(v/c)\cos\theta_{0}]\sigma(\nu)dt.
\end{equation}

\noindent Note that Eq. (D.9) differs from the parallel mirror equation (D.4) in that radiation of 
frequency $\nu$ incident from any direction sees the same cross-section of the sphere while 
this is not the case with the mirror,

	Therefore, the momentum transferred in the $z$-direction in the laboratory frame in time $dt$ by these photons 
is, by (D9) and half of (D3),

\begin{equation}\label{D10}
 -vdtd\xi=J\sigma(\nu)dt\Delta P_{0}
 =n(\nu_{0})d\nu_{0}d\Omega_{0}c[1-(v/c)\cos\theta_{0}]\sigma(\nu)dt(h\nu_{0}/c)[\cos\theta_{0}-(v/c)].
\end{equation}	
 
\noindent As before, we express $d\nu_{0}$ and $n(\nu_{0})\nu_{0}$ in terms of $\nu$ and $\theta_{0}$ ($\sigma$ 
is already in terms of $\nu$) and integrate over all $\Omega _{0}$: 

\begin{eqnarray}\label{D11} 
-vd\xi &&=d\nu h\sigma(\nu)\int_{0}^{\pi}d\Omega_{0}\bigg\{ 
\nu n(\nu)[\cos\theta_{0}-(v/c)]+
(\nu n(\nu))'\nu(v/c)\cos^{2}\theta_{0}\bigg\}\nonumber\\
&&=-vd\nu(4\pi/3)(h/c)\sigma(\nu)[3\nu n(\nu)-\nu(\nu n(\nu))'].
\end{eqnarray}

\noindent This is 2/3 of the comparable expression (D6) for the mirror, with the 
area $A$ replaced by the cross-section $\sigma(\nu)$.

	The classically calculated cross-section (i.e., the total scattered energy/sec divided by the incident energy/sec-area) 
for an electromagnetic wave of wavelength $>>R$ is\cite{Jackson}

\begin{equation}\label{D12}
 \sigma(\nu)=\bigg(\frac{8\pi}{3}\bigg)\bigg(\frac{2\pi\nu}{c}\bigg)^{4}R^{6}\bigg[\frac{\epsilon -1}{\epsilon+2}\bigg]
 \rightarrow \bigg(\frac{8\pi}{3}\bigg)\bigg(\frac{2\pi\nu}{c}\bigg)^{4}R^{6}  
\end{equation}

\noindent where, for simplicity, we shall only use the limit of large $\epsilon$.  
In Eq. (D12), $\sigma$ has been averaged over 
incident polarizations and summed over scattered polarizations. 

	Putting (D1) for $n(\nu)$ and (D12) for $\sigma(\nu)$ into (D11) yields 
	
\begin{equation}\label{D13}
 d\xi=\bigg(\frac{8\pi}{3}\bigg)\bigg(\frac{\nu}{c}\bigg)^{3}\bigg(\frac{h\nu}{kT}\bigg)h
 \frac{e^{\beta h\nu}}{[e^{\beta h\nu}-1]^{2}}d\nu\sigma(\nu)=
 (2\pi)^{4}\bigg(\frac{8\pi}{3}\bigg)^{2}\bigg(\frac{\nu}{c}\bigg)^{7}
 \bigg(\frac{h\nu}{kT}\bigg)\bigg(\frac{h}{c}\bigg)R^{6} 
 \frac{e^{\beta h\nu}}{[e^{\beta h\nu}-1]^{2}}d\nu.
\end{equation}

\noindent We remark that, if $d\nu\sigma(\nu)$ in the first equation of (D13) 
is replaced by $\int_{0}^{\infty}d\nu\sigma(\nu)=\pi e^{2}/mc$, the sum rule for an 
individual oscillator\cite{Jackson2} of mass m and resonant frequency $\nu$, we obtain the 
value of $\xi$ for a single oscillator given in reference\cite{Hoye}. 

	Upon integrating {D13} over $\nu$ we obtain the viscosity coefficient  
	
\begin{equation}\label{D14}
 \xi=\bigg(\frac{8}{9\pi}\bigg)\bigg(\frac{kT}{\hbar c}\bigg)^{8}\hbar R^{6}
 \int_{0}^{\infty}dz z^{8}\frac{e^{z}}{[e^{z}-1]^{2}}=\frac{4(2\pi)^{7}}{135}\bigg(\frac{kT}{\hbar c}\bigg)^{8}\hbar R^{6}
 \end{equation}	

\noindent since the integral=$(2\pi)^{8}/60$ ($\approx 8!$).  This result is used in Sections IIB and 1VC.

\section {A Gravitational Proposal}\label{Appendix E}

	Diosi\cite{Diosi} suggested a gravitationally based CSL-type collapse model with 
the collapse rate $\sim G$. However, it  
effectively had $a\approx$ the proton size and therefore too large a proton excitation rate, a flaw 
corrected by Ghirardi, Grassi and Rimini\cite{GGR} who added the standard $a$ to the model. 

	Penrose\cite{others}, perhaps unwilling to commit to a nonfundamental parameter $a$ (however, see 
Pearle and Squires\cite{PearleSquires2} for a ``derivation" of $a$ in terms of fundamental constants in the context of a 
gravitationally based model) has a more modest proposal. His suggestion is that, 
when quantum theory describes an object as being in a state of two superposed locations, 
collapse of the state to one of those locations will take place in a time equal to $\hbar$ divided by the 
gravitational energy required to move two real copies of the object from a completely 
overlapping configuration to these two locations.  For example, consider a sphere of mass $M$ and 
radius $R$.  Since the gravitational energy of two such spheres displaced by a small distance $D<<R$ is

\[ U(D)=\frac{GM^{2}}{R}\bigg[-\frac{6}{5}+\frac{1}{2}\bigg(\frac{D}{R}\bigg)^{2}\bigg], 
\]

\noindent then the time it takes a quantum state of a sphere in a superposition of 
two states separated by the distance $D$ to collapse to one or the other state is 

\begin{equation}\label{E.1}
\tau_{c}=2\hbar R^{3}/GM^{2}D^{2}
\end{equation}

 This is a minimalist proposal, not a complete dynamical theory.  
For example, it is silent on how to treat the collapse of the state of a sphere in a continuous 
superposition of locations (i.e., the usual wavefunction description of the CM of a sphere).  
Nonetheless, we shall have the temerity to make what we regard as a reasonable extrapolation to that situation, 
in order to estimate the random walk entailed by this proposal. 

\subsection{Equilibrium CM Wavepacket Size For A Sphere }

	First, consider the qualitative argument given after Eq. (3.5), for the equilibrium size of a CM wavefunction, 
applied to the sphere.  A CM wavepacket of width $D$ expands a distance $\sim(\hbar/MD)\Delta t$ in time 
$\Delta t$ due to the Schr\"odinger evolution.  Now, assume that the collapse is linear, in the sense that, 
in time $\Delta t$, if 
the wavepacket width is $D$, collapse acting alone makes it 
contract to $D[1-(\Delta t/\tau_{c})]$, where $\tau_{c}$ is 
given by Eq. (E.1).  If $D=s$ is the equilibrium width of the wavepacket, then the Schr\"odinger expansion is 
compensated by the collapse contraction, yielding $s\Delta t/\tau_{c}\sim (\hbar/Ms)\Delta t$ or

 \begin{equation}\label{E.2}
s^{4}\sim \frac{\hbar^{2}R^{3}}{GM^{3}}
\end{equation}

\noindent Eq. (E.2) may be compared to 
the CSL result (3.5): 

\[ s^{4}\sim \frac{\hbar a^{2}m^{2}}{\lambda M^{3}f(R/a)}.
\]

We may therefore regard this proposal's equilibrium CM  wavepacket size as giving the CSL size if  

\begin{equation}\label{e.3}
\lambda f(R/a)\sim \frac{Gm^{2}}{a\hbar}\bigg(\frac{a}{R}\bigg)^{3}
\end{equation}

\noindent In particular, if $R\sim a$ (and so $f\approx 1$), 

\begin{equation}\label{E.4}
\lambda \sim \frac{Gm^{2}}{a\hbar}\approx 10^{-23}{\rm sec}^{-1}
\end{equation}

\subsection{Translational Diffusion Of A Sphere}

	We may obtain the same result, that this proposal gives the CSL behavior for objects of size 
$\approx a$ with $\lambda$ having numerical value (E.4), from other considerations such as random walk of the sphere. 
In this case, the Schr\"odinger equation tells us that 
$d^{3}\overline {{{\langle {Q^{j}}^{2}\rangle}}}/dt^{3}=(2/M^{2})d\overline {{{\langle {P^{j}}^{2}\rangle}}}/dt$ (Eqs. (A7)).
The collapse, acting on a wavefunction of width $D$, narrows the wavefunction and, in so doing, increases the energy.
As we have remarked in the previous subsection, 

 \[ dD/dt=-D/\tau_{c}=-GM^{2}D^{2}/2\hbar R^{3}.
 \] 
 
\noindent From the uncertainty principle, $\overline{{{\langle {P^{j}}^{2}\rangle}}}\approx (\hbar/D)^{2}$, so 

\[ d\overline{{{\langle {P^{j}}^{2}\rangle}}}/dt\sim -(\hbar^{2}/D^{3})dD/dt\sim GM^{2}\hbar/R^{3}
 \] 
 
 \noindent (notice that the result is independent of $D$, as in CSL) and so 
 
\begin{equation}\label{E.5}
d^{3}\overline {{{\langle {Q^{j}}^{2}\rangle}}}/dt^{3}\sim G\hbar/R^{3}
\end{equation} 

\noindent (notice that the result is independent of $M$, as in CSL).  Comparison of Eq. (E.5) 
with the CSL result (3.4):

\begin{equation}\label{E.6}
 d^{3}\overline{{{\langle {Q^{j}}^{2}\rangle}}}/dt^{3}={\lambda\hbar^{2}f(R/a)\over m^{2}a^{2}} 
\end{equation}

\noindent yields the same ``effective" $\lambda$ given in (E.3)

\subsection{Rotational Diffusion Of A Disc}

	Angular random walk of a disc proceeds along the same lines.  For a thin disc of mass $M$ and radius $L$, 
the gravitational energy required to rotate one such disc through a small angle $\theta$ with respect to a second initially 
completely overlapping disc is $\sim (GM^{2}/L)\theta^{2}$ so

\begin{equation}\label{E.7}
\tau_{c}\sim \hbar L/GM^{2}\theta ^{2}.
\end{equation}

\noindent Here we utilize, from Eqs. (C.3), 
$d^{3}\overline {{{\langle {\theta}^{2}\rangle}}}/dt^{3}=(2/I^{2})d\overline {{{\langle {\cal L}^{2}\rangle}}}/dt$. 
According to this gravitational proposal, $d\theta/dt=-\theta/\tau_{c}$. From the uncertainty principle, 
$\overline {{{\langle {\cal L}^{2}\rangle}}}\sim (\hbar/\theta)^{2}$ so

\[d\overline {{{\langle {\cal L}^{2}\rangle}}}/dt\sim -\hbar^{2} /\theta^{3}d\theta/dt\sim \hbar^{2}/\theta^{2}\tau_{c}  
\]

\noindent and so

\begin{equation}\label{E.8}
d^{3}\overline {{{\langle {\theta}^{2}\rangle}}}/dt^{3}\sim \hbar^{2}/I^{2}\theta^{2}\tau_{c}\sim G\hbar/L^{5}.
\end{equation}

\noindent Eq. (E.8) may be compared with the CSL result (6.5) 

\begin{equation}\label{E.9}
d^{3}\overline {{{\langle {\theta}^{2}\rangle}}}/dt^{3}\sim \lambda (\hbar/ma^{2})^{2}f_{ROT}(L/2a)
\end{equation}

\noindent which yields the ``effective" $\lambda$ 

\begin{equation}\label{E.10}
\lambda f_{ROT}(L/2a)\sim \frac{Gm^{2}}{a\hbar}\bigg(\frac{a}{L}\bigg)^{5}
\end{equation}

\noindent In our proposed experiment, for which $L\sim a$ (and so $f_{ROT}(L/2a)\approx 1$),  
the ``effective" $\lambda$  is again given by (E.4). 

The results obtained here are effectively the same as would be obtained with the modified Diosi model.  

\section {Thermal Source Of The Fluctuations?}\label{Appendix F}

	It is fun to speculate that the collapse-inducing fluctuations of $w$ may come from a thermal 
bath, as do so many other fluctuations in physics. 

	Since a thermal bath defines a preferred reference frame (i.e., the frame in which the bath 
medium has zero average momentum density), this would preclude a truly special relativistically 
invariant collapse model.  But, anyway, the universe is not truly special relativistically invariant, 
possessing as it does the preferred comoving reference frame. Moreover, this reference frame is endowed with 
the 2.7$^{\circ}$K thermal radiation bath. So, one might entertain the idea that the 
fluctuations of $w$ arise from some unspecified medium in thermal equilibrium with the 2.7$^{\circ}$K radiation.  

	For an object in random walk, we note that the $\sim t^{3}$ time dependence of 
$\overline {{{\langle {Q^{j}}^{2}\rangle}}}$ given by Eq. (3.4) for CSL is also the 
time dependence of $(\Delta x)^{2}$ given by Eq. (2.7c) for ordinary Brownian motion when 
$t$ is very much smaller than $\tau=\xi/M$ (which characterizes the time scale of the 
approach to thermal equilibrium).  Since objects show no sign of reaching thermal equilibrium today, 
we may assume that $\tau$ is larger than the age of the universe, 
$\tau\equiv\gamma 50 \lambda_{CSL}^{-1}$ (since $\lambda_{CSL}^{-1}=10^{16}$sec$\approx3\cdot 10^{8}$yr) 
with $\gamma>1$.  

	Continuing in the same lighthearted vein, we propose that, for a fundamental object, the nucleon, 
the two sources of the $\sim t^{3}$ behavior, thermal and CSL,  
may be identified, and we equate Eqs. (2.7a) and (3.4), obtaining 

\begin{equation}\label{F.1}
  \frac{kT}{m\tau}\approx \frac{\lambda_{CSL}\hbar^{2}}{m^{2}a_{CSL}^{2}} \thinspace\thinspace
\thinspace\thinspace\thinspace\thinspace{\rm or}\thinspace\thinspace
\thinspace\thinspace\thinspace\thinspace
KT\approx 50\gamma \frac{\hbar^{2}}{m^{2}a_{CSL}^{2}} 
\end{equation}

\noindent (we have set $M=m$ and $f(R/a)=1$). 

	When $T=2.7^{\circ}$K then $kT\approx 2.5\cdot 10^{-4}$eV.  
The energy $\hbar^{2}/m^{2}a_{CSL}^{2}\approx 4\cdot 10^{-9}$eV. Thus, Eq. (F.1) 
implies $\gamma\approx 10^{3}$, which is consistent.  

	Of course, in the speculation above there is no need to choose $(\lambda^{-1}, a)$ to have their 
CSL numerical values.  The appropriate generalization of (8.16) is

\begin{equation}\label{F.2}
  \frac{\lambda^{-1}}{\lambda_{CSL}^{-1}}\bigg(\frac{a}{a_{CSL}}\bigg)^{2}\approx \frac{\gamma}{10^{3}} \thinspace\thinspace
\thinspace\thinspace\thinspace\thinspace{\rm or}\thinspace\thinspace
\thinspace\thinspace\thinspace\thinspace
\lambda^{-1}a^{2}\approx10^{3}\gamma.
\end{equation}

There is quite a range of $\lambda$ and $a$ consistent with (F.2) and present constraints (see FIG. 2),   
especially in view of the flexibility in choosing $\gamma$.

\begin{figure}
\caption{A graph of $f_{ROT}(\alpha, \beta)$ as a function of $\alpha\equiv$(disc radius)$/2a$, for 
various values of $\beta\equiv$(disc thickness)$/2a$.}
\end{figure}

\begin{figure}
\caption{A graph of $\log_{10}\lambda^{-1}$ vs $\log_{10} a$ for various constraints: the 
constraint boundaries specified in Eqs. (8.1), (8.2), (8.3), (8.4) are respectively labelled 
1, 2, 3, 4.}
\end{figure}

\begin{table}
\caption{CSL diffusion in vacuum: rms distance $\Delta Q$cm for various 
sphere radii $R$ and times $t$.}
\begin{tabular}{cccc}
&\multicolumn{2}{c}{$t$ in sec}\\
$R$ in cm&$10$&$10^{3}$&$10^{5}$\\
\tableline
$10^{-6}$&$8\cdot 10^{-6}$&$8\cdot 10^{-3}$&8\\
$10^{-5}$&$6\cdot 10^{-6}$&$6\cdot 10^{-3}$&6\\
$10^{-4}$&$2\cdot 10^{-7}$&$2\cdot 10^{-4}$&$2\cdot 10^{-1}$\\
$10^{-2}$&$6\cdot 10^{-11}$&$2\cdot 10^{-8}$&$2\cdot 10^{-5}$\\
$1$&$6\cdot 10^{-15}$&$2\cdot 10^{-12}$&$2\cdot 10^{-9}$\\
\end{tabular}
\end{table}

\begin{table}
\caption{CSL rms equilibrium center of mass 
wavefunction size $s_{\infty}$ and characteristic time $\tau_{s}$ 
to reach that size in vacuum for various radii $R$ of a sphere of density 1gm/cc.}
\begin{tabular}{ccc}
$R$ in cm&$s_{\infty}$ in cm&$\tau_{s}$ in sec\\
\tableline
$10^{-6}$&$7\cdot 10^{-5}$&20\\
$10^{-5}$&$4\cdot 10^{-7}$&.6\\
$10^{-4}$&$1\cdot 10^{-8}$&.6\\
$10^{-2}$&$4\cdot 10^{-11}$&6\\
$1$&$1\cdot 10^{-13}$&60\\
\end{tabular}
\end{table}


\begin{references}

\bibitem{Schrodinger}E. Schr\"odinger, Die Naturwissenschaften 23, 807 (1935).

\bibitem{PearleCSL} P. Pearle, Phys. Rev. A {\bf 39}, 2277 (1989)

\bibitem{GPR}  G. C. Ghirardi, P. Pearle and A. 
Rimini, Phys. Rev. A {\bf 42}, 78 (1990).

\bibitem{GRW} G. C. Ghirardi, A. Rimini and T. Weber, Phys. Rev. D {\bf 34}, 470 (1986); 
Phys. Rev. D {\bf 36}, 3287 (1987); Found. Phys. {\bf 18}, 1, (1988). 

\bibitem{Pearle} P. Pearle, Phys. Rev. D {\bf 13}, 857 (1976); 
Int'l. Journ. Theor. Phys. {\bf 48}, 489 (1979); 
Found. Phys. {\bf 12}, 249 (1982); 
Phys. Rev. D {\bf 29}, 235 (1984); 
in {\it The Wave-Particle Dualism}, edited by S. Diner et. al (Reidel, Dordrecht 1984); 
Journ. Stat. Phys. {\bf 41}, 719 (1985); 
in {\it Quantum Concepts in Space and 
Time}, edited by R. Penrose and C. J. 
Isham (Clarendon, Oxford, 1986); Phys. Rev. D {\bf 33}, 2240 (1986); 
in {\it New Techniques in Quantum Measurement Theory}, 
edited by D. M. Greenberger (N.Y. Acad. of Sci., N.Y., 1986), p.539. 

\bibitem{others} For some other collapse models, see I. C. Percival, Proc. Roy. Soc. A {\bf 451}, 503 (1995) and
{\it Quantum State Diffusion}, (Cambridge Univ. Press, Cambridge, 1998);
L. P. Hughston, Proc. Roy. Soc. A {\bf 452}, 953 (1995); R. Penrose, Gen. Rel. and Grav. {\bf 28}, 581 (1996);
S. L. Adler and L. P. Horwitz, Journ. Math. Phys. {\bf 41}, 2485 (2000); D. Fivel, Phys. Rev. A {\bf 56}, 146 (1997).

\bibitem{PearleNaples} For a recent review of CSL, see P. Pearle in 
{\it Open Systems and Measurement in Relativistic Quantum theory}, 
edited by A. Miller (Plenum, New  York 1990), p.167.
 
\bibitem{GR} G. C. Ghirardi and A. Rimini in {\it Sixty-Two Years of Uncertainty}, 
edited by H.P. Breuer and F, Petruccione (Springer, Heidelberg 1999), p.195.

\bibitem{Squires} E. J. Squires, Phys. Lett. A {\bf 158}, 431 (1991).

\bibitem{Ballentine} L. E. Ballentine, Phys. Rev. A {\bf 43}, 9 (1991).

\bibitem{PearleSquires}	P. Pearle and E. Squires, Phys. Rev. Lett. {\bf 73}, 1 (1994).

\bibitem{Collett} B. Collett, P. Pearle, F. Avignone and S. Nussinov, 
Found. Phys. {\bf 25}, 1399 (1995).     

\bibitem{Ring} P. Pearle, James Ring, J. I. Collar and F. T.  Avignone III,
 Found. Phys. {\bf 29}, 465 (1999). 

\bibitem{Karolyhazy} F. Karolyhazy, Nuovo Cimento {\bf42A}, 1506 (1966); 
F. Karolyhazy, A Frenkel and B. Lukacs in 
{\it Physics as Natural Philosophy}, edited by A. 
Shimony and H. Feshbach (M.I.T. Press, Cambridge 1982), p. 204; 
in {\it Quantum Concepts in Space and Time}, edited by R. Penrose and 
C. J. Isham (Clarendon, Oxford 1986), p. 109; A. Frenkel, Found. Phys. {\bf 20}, 159 (1990). 

\bibitem{Gabrielse} G. Gabrielse et. al, Phys. Rev. Lett. {\bf 65}, 1317 (1990).   
 
\bibitem{Mazo} For a nice treatment see R. M. Mazo in {\it Stochastic Processes in Nonequilibrium Systems, 
Lecture Notes in Physics {\bf 84}}, edited by L. Garrido, P. Seglar and P. J. Shepard (Springer-Verlag, Berlin 1978), p. 53. 

\bibitem{Millikan} R. A. Millikan, Phys. Rev. {\bf 32}, 349 (1911); Phys. Rev. {\bf 22}, 1 (1923). 

\bibitem{Aerosol} M. D. Allen and O. G. Raabe, Aerosol Sci. and Tech. {\bf 4}, 269 (1985). 

\bibitem{Cunningham} E. Cunningham, Proc. Roy. Soc. {\bf 83}, 357 (1910). 

\bibitem{Epstein} P. S. Epstein, Phys. Rev. {\bf 23}, 710 (1924).   

\bibitem{EinsteinandHopf} A. Einstein and L. Hopf, Ann. der Phys. {\bf 33}, 1105 (1910).

\bibitem{Einstein} A. Einstein, Phys. Zeit. {\bf 10}, 185 (1909). 

\bibitem{Einstein1904} A. Einstein, Ann. der Phys. {\bf 17}, 549 (1905).
 
\bibitem{BGG} F. Benatti, G. C. Ghirardi and R. Grassi, Found. Phys. {\bf 35}, 5 (1995).  
 
\bibitem{BG} A. Bassi and G. C. Ghirardi, Brit. J. Phil. Sci. {\bf 50}, 719 (1999). 

\bibitem{Diosi4} L. Diosi, Phys. Lett. {\bf A132}, 233 (1988).

\bibitem{Andrade} E. N. daC. Andrade and R. C. Parker, Proc. Roy. Soc. {\bf 159}, 507 (1937). 

\bibitem{Lamb} H. Lamb, {\it Hydrodynamics} (Dover, N.Y. 1945), p. 605.

\bibitem{Einstein Rot} A. Einstein, Ann. der Phys. {\bf 19}, 371 (1906). 

\bibitem{Lamb2} H. Lamb, Op. Cit. p. 589. 

\bibitem{Pearleenergy} P. Pearle, Found. Phys. {\bf 30}, 1145 (2000).

\bibitem{AvignoneRing}  We are indebted to Frank Avignone for supplying recent data and 
to Jim Ring for analyzing it. 

\bibitem{QFu} Q. Fu. Phys. Rev. A56, 1806 (1997).

\bibitem{Diosi}L. Diosi, Phys. Rev. A{\bf 40}, 1165 (1989).

\bibitem{GGR}G. C. Ghirardi, R. Grassi and A. Rimini, Phys. Rev. A {\bf 42}, 1057 (1990).

\bibitem{PearleSquires2} P. Pearle and E. Squires, Found. Phys. {\bf 26}, 291 (1996).

\bibitem{ABGG} F. Aicardi, A. Borsellino, G. C. Ghirardi and R. Grassi, Found. Phys. Lett {\bf 4}, 109 (1991).

\bibitem{Chen} Chen et. al., J. Aerosol. Sci. {\bf 24},181 (1993).            
 
\bibitem{Tannenbaum} Tannenbaum et. al. J. Vac. Sci. submitted. Cornell Project 789-99.

\bibitem{Paul} Paul Rev. Mod. Phys. {\bf 60} ,531 (1990).

\bibitem{Wuerker} Wuerker at. al. J. Applied. Phys. {\bf 30}, 342 (1958).

\bibitem{Arnold1} S. Arnold, J.H. Li, S. Holler, A. Korn and A.F. Izmailov, J. Appl. Phys. {\bf 78}, 3566 (1995).

\bibitem{Arnold2} S. Arnold, L. M. Foley and A. Korn, J. Appl. Phys. {\bf 74}, 4291 (1993).

\bibitem{Hoye} J. S. Hoye and I Brevik, Physica A {\bf 196}, 241 (1993).  We would like to thank 
Peter Milloni for calling our attention to this paper.

\bibitem{Jackson} J. D. Jackson, {\it Classical Electrodynamics} (Wiley, N.Y. 1975), p. 414.            

\bibitem{Jackson2} J. D. Jackson, Ibid p. 805.   

\end{references}
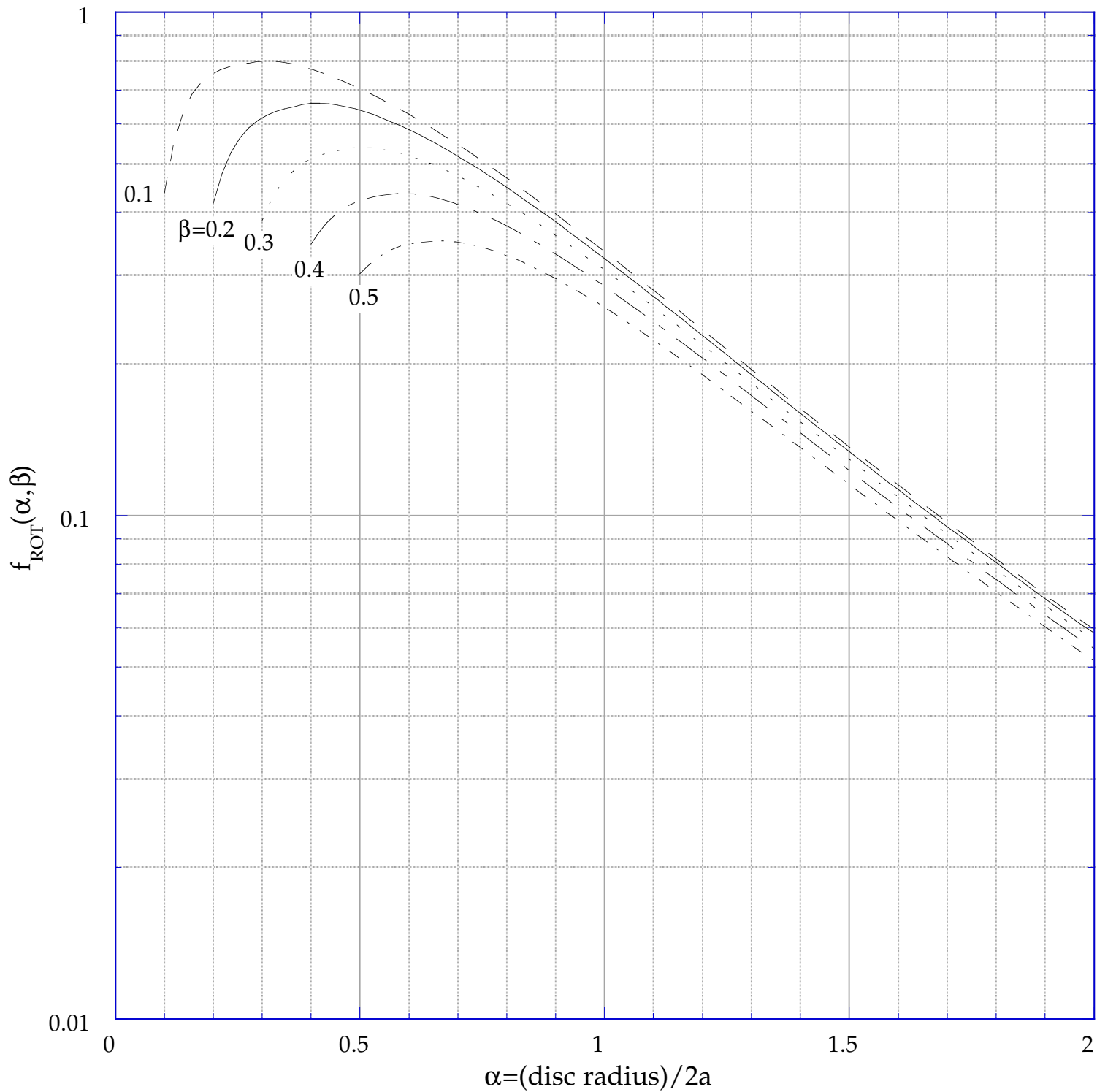
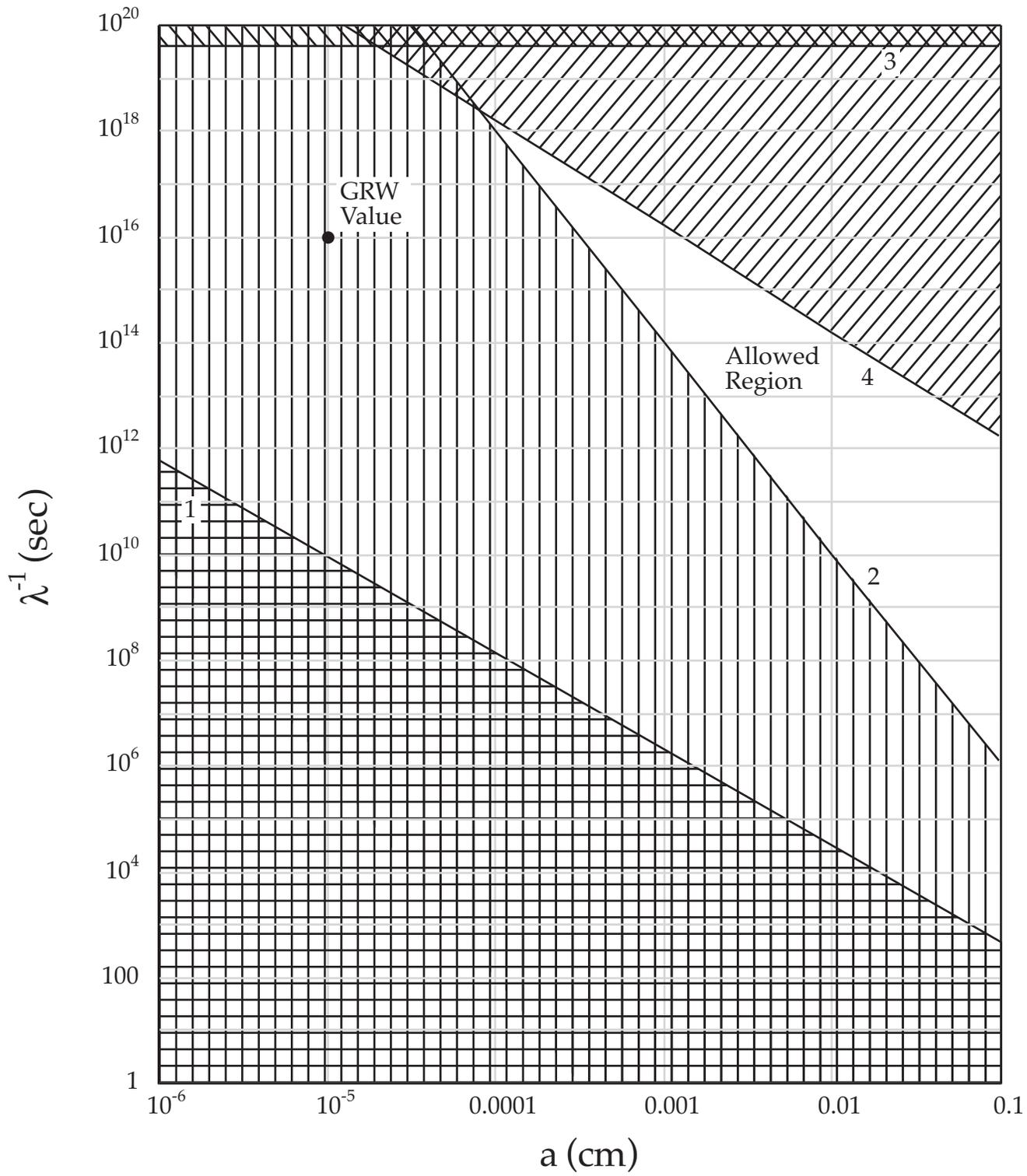
\end{document}